\documentclass[fleqn,usenatbib]{mnras}
\usepackage{newtxtext,newtxmath}
\usepackage[T1]{fontenc}

\DeclareRobustCommand{\VAN}[3]{#2}
\let\VANthebibliography\thebibliography
\def\thebibliography{\DeclareRobustCommand{\VAN}[3]{##3}\VANthebibliography}

\usepackage{booktabs}

\usepackage{graphicx}	
\usepackage{amsmath}	
\usepackage{arydshln}
\usepackage{ulem}

\title[Tracing star formation in the Rosette Nebula]{Tracing star formation across the Rosette nebula using deep Spitzer imaging}

\author[D. Capela et al.]{D. Capela,$^{1}$\thanks{E-mail: dc295@st-andrews.ac.uk}
A. Scholz,$^{1}$
K. Mužić,$^{2}$
H. Bouy,$^{3}$
B. Damian,$^{1}$
V. Almendros-Abad,$^{4}$
and A. Bayo$^{5}$
\\
$^{1}$SUPA, School of Physics \& Astronomy, University of St Andrews, North Haugh, St Andrews KY16 9SS, UK\\
$^{2}$Instituto de Astrofísica e Ciências do Espaço, Faculdade de Ciências, Universidade de Lisboa, Ed. C8, Campo Grande, 1749-016 Lisbon, Portugal\\
$^{3}$Laboratoire d’Astrophysique de Bordeaux, Univ. Bordeaux, CNRS, B18N, Allée Geoffroy Saint-Hillaire, 33615 Pessac, France\\
$^{4}$Istituto Nazionale di Astrofisica (INAF) – Osservatorio Astronomico di Palermo, Piazza del Parlamento 1, 90134 Palermo, Italy\\
$^{5}$European Southern Observatory, Karl-Schwarzschild-Strasse 2, D-85748 Garching bei München, Germany}

\date{Accepted 2026 September 19. Received 2026 September 04; in original form 2026 April 29}

\pubyear{\the\year{}}

\begin{document}
\label{firstpage}
\pagerange{\pageref{firstpage}--\pageref{lastpage}}
\maketitle

\begin{abstract}
We present a deep large-scale infrared survey of young stellar objects in the Rosette Nebula. Our analysis is based on stacked archival Spitzer/IRAC images. The images have been treated with the software DeNeb to remove nebulous emission. Our catalogue contains $\sim$20000 sources with photometry in all 4 IRAC bands from 3.6 to 8.0$\,\mu m$, complemented with near-infrared and optical data. From this database, we identify YSOs using multi-filter colour-colour criteria. We select 1528 YSOs, from which 1303 are Class II and 225 are Class I. Using this sample, we compile a list of clusters and groups of YSOs in the Rosette region, including previously known ones, plus three newly identified groups. Using our census of YSOs, in combination with the literature, we derive the Class I ratios and disk fractions for clusters in the nebula. The main clusters in the central cavity, NGC~2244 and NGC~2237 are almost devoid of Class I sources, indicating that star formation has ceased in this area. On the other hand, in an annulus around the centre with radii of 11-16\,pc, roughly coinciding with the expanding HII front, about 30\% of the YSOs are Class I, a clear sign of ongoing star formation. Other groups further out in the cloud also exhibit a high Class I ratio. Similar patterns are seen in the disk fractions. Our results are consistent with possible triggering by the expanding HII region, combined with star formation beyond the ionisation front.
\end{abstract}




\section{Introduction}
\label{sec:intro}

Star clusters serve as the primary sources of both stellar and substellar objects in the Universe \citep{Lada2003,Portegies}. Within young star clusters, objects spanning at least four orders of magnitude in mass are born, ranging from high-mass stars, many times the mass of our Sun, to those of lower mass, similar to the Sun itself, and extending to substellar objects like brown dwarfs and planetary-mass objects. Historically, young stellar objects are grouped in classes based on the appearance of their infrared spectral energy distribution, which broadly links to evolutionary state: Class I are embedded sources dominated by infrared light from their surrounding envelopes, Class II are visible in the optical, with infrared excess from circumstellar disks, and Class III have spectral energy distribution consistent with diskless stars, with little to no infrared excess \citep{schulz}.

\begin{figure}
    \includegraphics[width=\columnwidth]{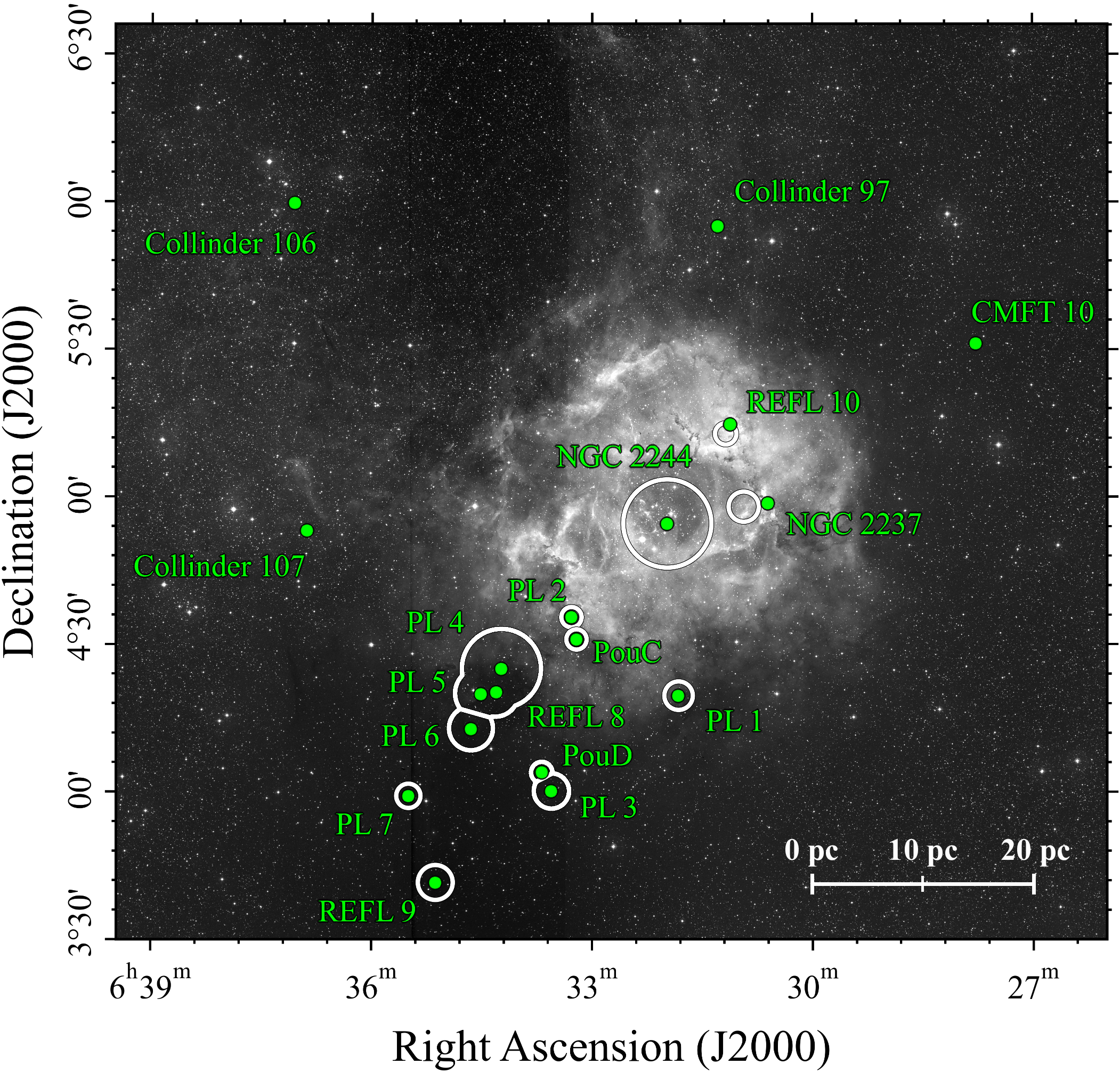}
    \caption{Digitized Sky Survey (DSS) image of the Rosette Nebula region, showing the positions of previously known stellar groups (green circles) along with their corresponding names. The PL, REFL, Pou, and CMFT designations are from \citet{PL7, Roman-Zuniga, Poulton, Cambresy}, respectively. Additionally, the white areas indicate the regions used in Section \ref{sec:stellargroups} to represent the positions occupied by the members of these known stellar groups. A parsec scale for a distance of 1500~pc is shown in the bottom right.}
    \label{fig:RosNebClusters}
\end{figure}

One of the prominent star forming regions in the northern sky is the Rosette Nebula, a massive star-forming cloud of gas and dust in the constellation Monoceros. The region is situated about 1500~pc from Earth and spans a diameter of approximately 40~pc \citep{2019Kuhn, Lim, Muzic_2022}. The nebula has a characteristic structure, with a large rosette-shaped HII region -- hence the name. At the core of the Rosette Nebula lies the star cluster NGC~2244, comprising a large OB association, whose stellar winds are presumed to have evacuated most of the parent material from the centre of the HII bubble \citep{Roman2}. NGC~2244 is estimated to be around 2 Myr old \citep{Perez, Lim, Muzic_2022}. Besides this central cluster, the Rosette Nebula also hosts other stellar groups that were previously identified in the studies by \citet{PL7, Roman-Zuniga, Poulton, Cambresy}, see Fig.~\ref{fig:RosNebClusters}. Star formation within the Rosette Nebula may be related to both its early evolution and its ongoing interaction with the HII front \citep{Roman2}. 

This paper is the fourth in a series of studies by our group to uncover the stellar and substellar content of the Rosette region. In \citet{Muzic2019}, we estimate the membership of the central cluster NGC~2244, using deep near-infrared imaging and a statistical approach to remove contamination. This leads to the first Initial Mass Function (IMF) for this region which includes the substellar domain, with a cutoff at 0.02$\,M_{\odot}$. In \citet{Muzic_2022}, the entire nebula is investigated, using archival optical and infrared imaging, plus Gaia astrometry, to select stellar members with a probabilistic random forest. This paper includes a characterisation of the whole region, a kinematic analysis, and a mass function for the stellar population. In \citet{Almendros2023}, we present an infrared spectroscopic survey of candidate very low mass (down to $\sim0.03 M_{\odot}$) members of NGC~2244, which includes the identification of the first spectroscopically confirmed brown dwarfs in this cluster.

In this paper, we will use the deep Spitzer catalogue, introduced in \citet{Almendros2023}, to study the population of objects with infrared excess, for a large portion of the Rosette Nebula.  In Section \ref{sec:dataset} we describe the dataset and catalogues used for our analysis. We use those catalogues to select young stellar objects (YSOs), described in depth in Section \ref{sec:selection}. In Section \ref{sec:Sec4}, we discuss the spatial distribution of the selected sample. In Section \ref{sec:population}, we examine the evolutionary stage of each stellar group. Our conclusions are presented in Section \ref{sec:conclusion}.

\section{Dataset}
\label{sec:dataset}

\subsection{Mid-infrared photometry}

The foundation for this paper is the Spitzer mid-infrared (MIR) catalogue introduced by \citet{Almendros2023}. The catalogue comprises deep photometry from IRAC \citep[InfraRed Array Camera;][]{IRAC} on the Spitzer Space Telescope \citep{Spitzer}. IRAC is a four channel photometer with bands at central wavelengths of 3.6, 4.5, 5.8 and 8.0$\,\mu  m$, commonly referred to as IRAC1, IRAC2, IRAC3 and IRAC4, respectively. The Rosette region was observed by Spitzer/IRAC several times; our catalogue compiles all existing datasets. Specifically, we make use of data from programs 30726, 3394, 37, 40359, 61071, 61073. The data is from a total of 9 epochs, ranging from 2004 to 2011.  The observational details are also summarised in \citet{Almendros2023}. The exact parameters for these Spitzer programs can be retrieved from the Spitzer Heritage Archive.

\begin{figure}
    \includegraphics[width=\columnwidth]{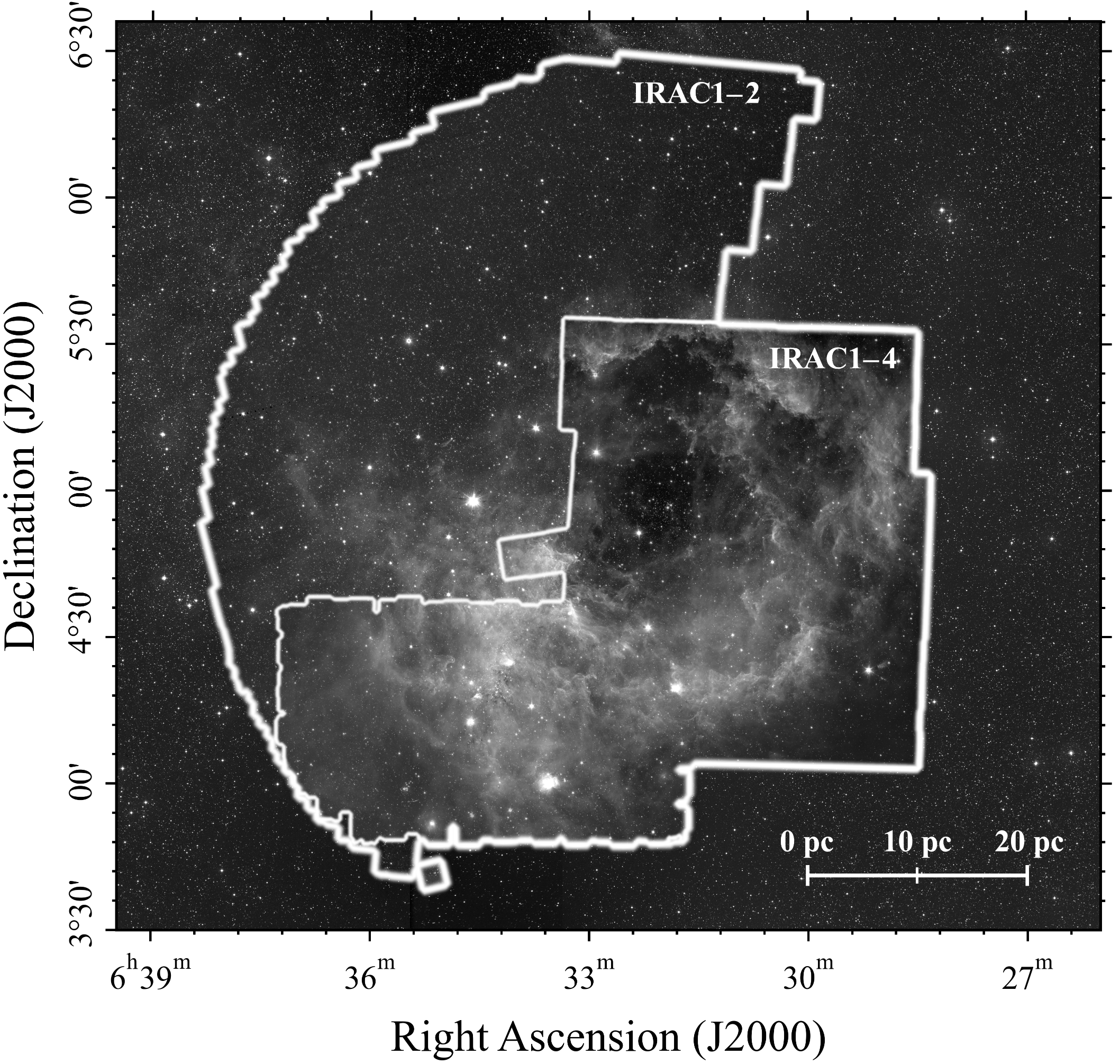}
    \caption{Overplotted onto the DSS background image of the Rosette Nebula region is the part of the IRAC2 mosaic indicating where we have IRAC1 and IRAC2 coverage. Overlaid on this, with 50\% opacity, is the region where photometric data in all IRAC channels are available, shown as part of the IRAC3 mosaic. Small gaps and artifacts were interpolated for clarity. A parsec scale for a distance of 1500~pc is shown in the bottom right.}
    \label{fig:Rosette Nebula}
\end{figure}

The MIR data was retrieved from the Spitzer Heritage Archive within a $1.5^\circ$ radius centred on NGC~2244 ($\alpha = 06{:}31{:}58.51$, $\delta = +04{:}54{:}35.7$). We started with the CBCD version (Corrected Basic Calibrated Data), and combined these images into mosaics, one per band, using {\it MOPEX} (MOsaicker and Point source EXtractor; \citealp{Mopex}), following recommended routines. For this study, we only used the long exposure frames with 10.4\,s exposure time. The coverage maps reveal that the exposure time is mostly uniform across the field, except for a portion of the core containing NGC~2244, which was observed with greater depth. We show the Spitzer fields overplotted onto a DSS image of the Rosette region in Fig.~\ref{fig:Rosette Nebula}.

\begin{figure}
    \includegraphics[width=\columnwidth]{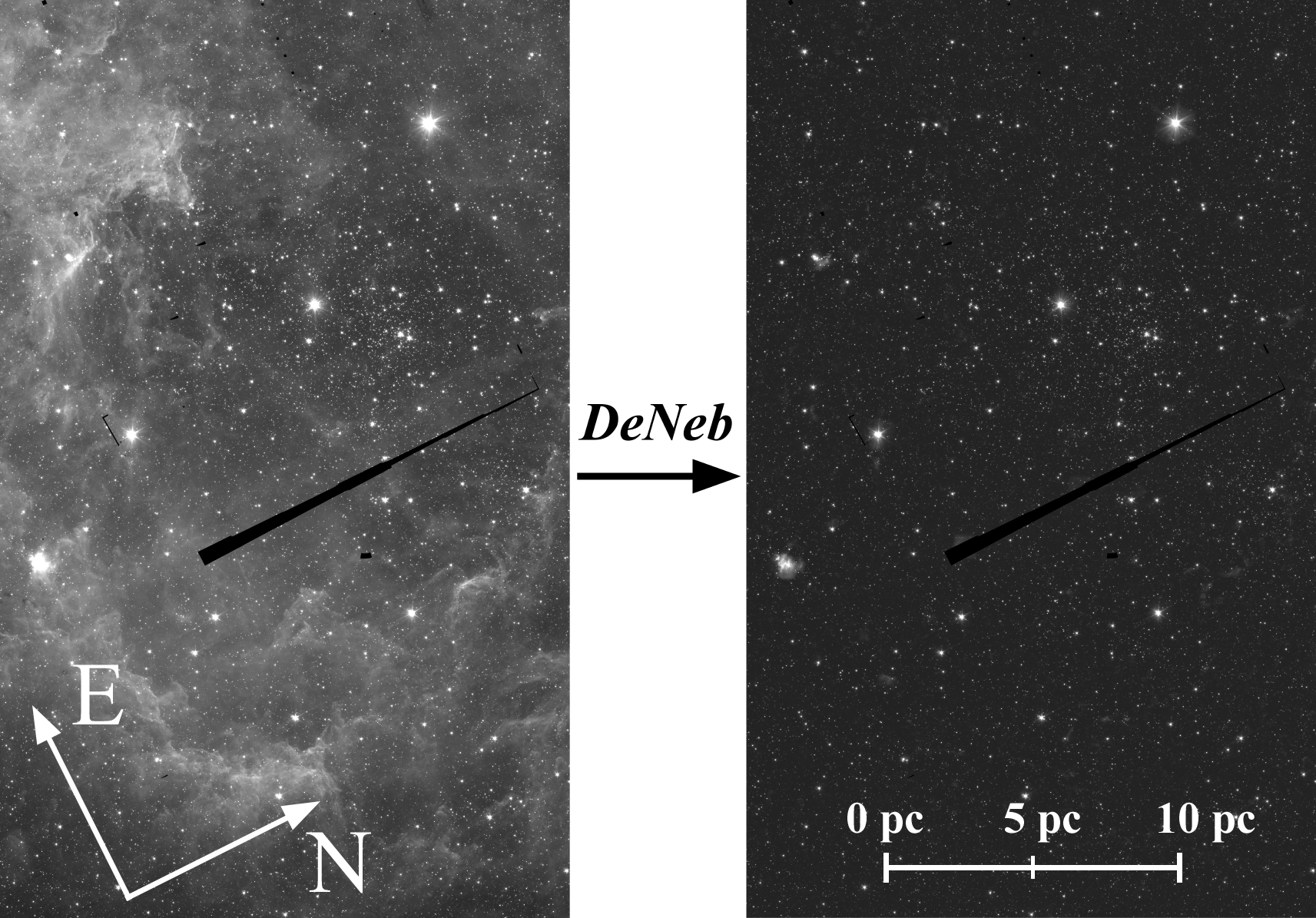}
    \caption{A portion of the IRAC1 mosaic showing the original image, including diffuse nebula emissions (left panel), and the resulting mosaic after DeNeb processing (right panel), on which aperture photometry was performed. The orientation of the image is depicted on the left and a parsec scale for a distance of 1500~pc is shown in the bottom right.}
    \label{fig:DeNeb_image}
\end{figure}

On the mosaics, we applied the {\it DeNeb} algorithm, described in detail in Bertin et al. (in prep.) and used previously in \citet{Almendros2023}, \citet{Bouy_2025}, \citet{Afonso}. In short, DeNeb separates the emissions from point sources and the background emission, respectively. The software achieves that using a convolutional neural network model, that has been trained on images of pure nebula. It has been demonstrated that images treated with DeNeb enhance the sensitivity for faint point sources significantly. For more information and an evaluation of performance of the software, we refer to \citet{Afonso}. Fig.~\ref{fig:DeNeb_image} shows a portion of an IRAC1 mosaic to demonstrate the results of applying DeNeb to the image.

We created a source catalogue on the cleaned mosaics using {\it SExtractor} \citep{SExtractor}, with the goal to maximise completeness, including faint sources. This necessarily means that the catalogues also contain artefacts and contamination. Specifically, we used a detection threshold of 2.5$\sigma$, and required that detections cover a minimum area of 5 pixels. The detections were filtered with a standard 'mexhat' filter with a size of $9\times 9$ pixels.  For deblending, we used a threshold of 32 and a minimum contrast of $10^{-7}$.

\begin{figure}
    \includegraphics[width=\columnwidth]{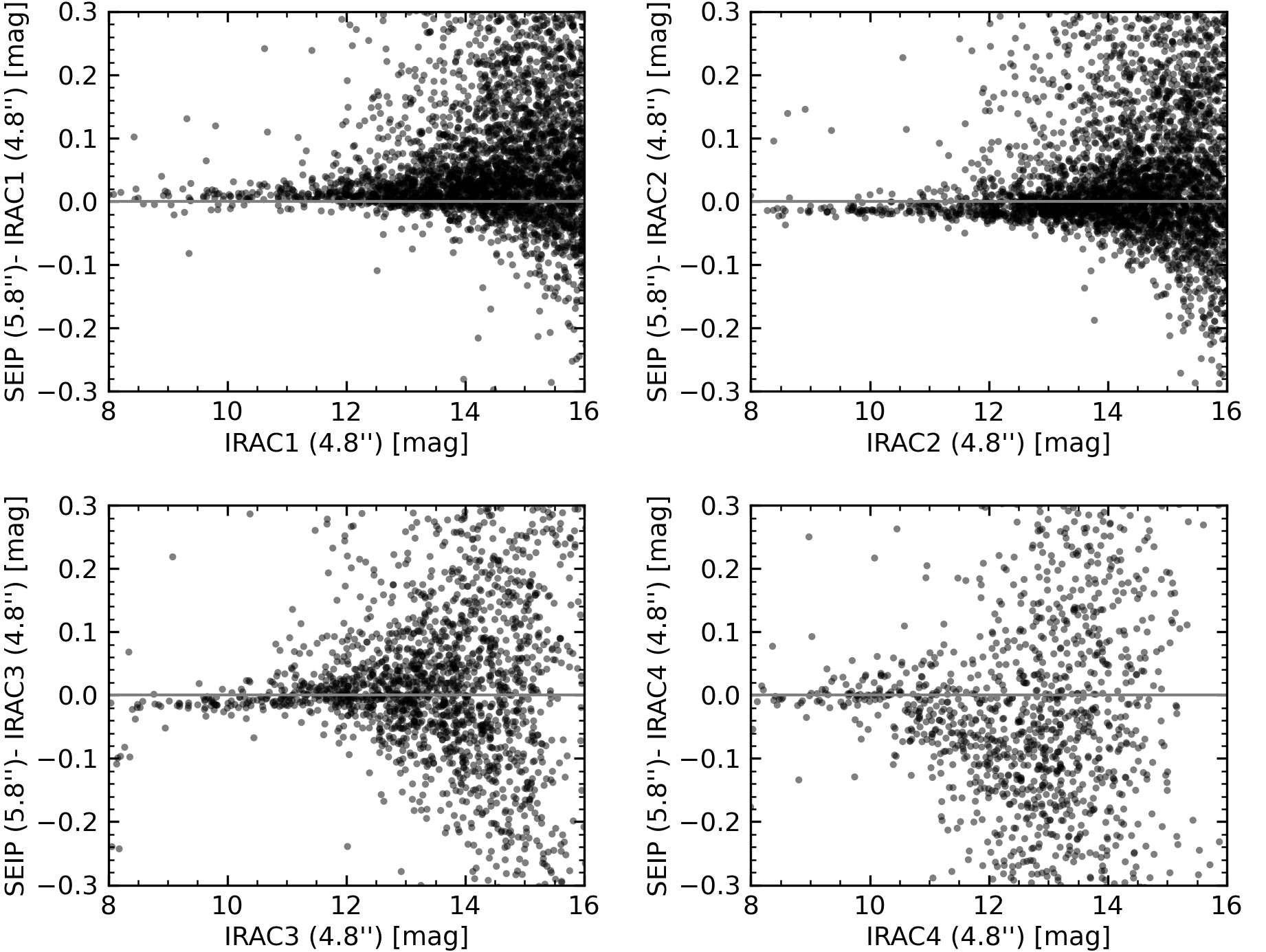}
    \caption{Photometric comparison between the magnitudes of sources measured in the DeNeb IRAC mosaic with a 4.8" aperture and their difference from the SEIP 5.8" aperture for each IRAC filter. The sample comprises sources extracted from SEIP data within a 15$'$ radius of NGC~2244’s centre.}
    \label{fig:DeNeb}
\end{figure}

We carried out aperture photometry with the {\it APEX} tool, using three different apertures of 2.4, 3.6 and 4.8" (corresponding to 4, 6, 8 of the native 0.6" pixels) and a background annulus from 14.4 to 24" (or 24 to 40 pixels). The recommended aperture corrections were applied to the fluxes. To validate the photometry we compared with publicly available data from SEIP \citep[Spitzer Enhanced Imaging Products;][]{IRSA}. For the smallest aperture and the longest wavelength, we see a systematic offset between our magnitudes and those from SEIP. For the largest apertures, the measurements are very well aligned, as demonstrated in Fig.~\ref{fig:DeNeb}. The photometry derived using the largest aperture was adopted throughout this work.

We merged the data from the four IRAC filters, using a 1" matching radius. This value was chosen because it approximates the spatial resolution of IRAC. The cross-matching of the four catalogues was performed simultaneously using a friends-of-friends algorithm. All detections were pooled into a single list, and any two detections separated by less than 1" were linked. Sources connected in this way were grouped together. Where a band contributed multiple candidates to a group, the detection minimising the overall positional dispersion of the group was retained. The final source position was taken as the mean RA/Dec of the selected members. Sources detected in only one single IRAC band were excluded, resulting in an IRAC catalogue containing a total of 392,982 sources. On average, all matched detections fall within 0.21" of the group mean position.

\subsection{Optical to near-infrared photometry}

To complement the IRAC photometry with data in near-infrared (NIR) and optical bands, we matched our catalogues with archival data from UKIDSS, 2MASS, and Pan-STARRS.

We retrieved a $1.5^\circ$ photometric dataset centred on NGC~2244 from the UKIDSS (UKIRT Infrared Deep Sky Survey) DR11 Galactic Plane Survey \citep{UKIDSS_DR11}. The catalogue provides photometry in the $J$-, $H$- and $K$-bands. We first removed objects likely to be detector noise or galaxies by retaining only sources classified as stars or probable stars (\textit{mergedClass} equal to $-1$ or $-2$). Duplicate detections were removed by retaining only primary sources (\textit{priOrSec} $= 0$ or \textit{priOrSec} $=$ \textit{frameSetID}). Detections were additionally required to satisfy a point-source probability exceeding 90\% (\textit{pStar} > 0.9), and sources affected by saturation or other processing artefacts were excluded by imposing \textit{ppErrBits} < 256 in each band.

Subsequently, we used the corrections outlined in \citet{Hodgkin2009} to address the systematically underestimated uncertainties present in the UKIDSS catalogue. We then employed the system of colour-correction equations from the same work to align our data with the 2MASS (Two Micron All-Sky Survey; \citealp{2MASS}) system, propagating the corrected uncertainties through the transformation.

We additionally retrieved a $1.5^\circ$ 2MASS all-sky point source catalogue centred on NGC~2244 \citep{2MASS}, which provides $J$-, $H$- and $K_{\mathrm{s}}$-band photometry. Photometry was kept if its quality flag, \textit{Qflg}, was C or higher quality, a valid aperture or profile-fit measurement (\textit{Rflg} $=$ 1, 2 or 3), a clean or properly deblended fit (\textit{Bflg} $> 0$), and not flagged as an artefact contamination or confusion (\textit{Cflg} $= 0$). We then removed sources blended with extended objects (\textit{Xflg} $= 0$) or matching known Solar System objects (\textit{Aflg} $= 0$), and kept only detections marked with the use source flag parameter (\textit{use} $= 1$).

We noted that when comparing 2MASS and UKIDSS magnitudes, an offset begins to appear for the brightest objects in the UKIDSS catalogue due to saturation effects. To avoid this issue, objects from the UKIDSS catalogue with K$_S$ < 11.5~mag were excluded.

Additionally, we retrieved Pan-STARRS (Panoramic Survey Telescope and Rapid Response System) Release 1 (PS1) Survey \citep{PanSTARRS} data from a $1.5^\circ$ circle centred on NGC~2244. This includes PSF photometric data in the $g$, $r$, $i$, $z$ and $y$ filters. We cleaned the catalogue using the \textit{Qual} bitmask, retaining only sources flagged as having a good-quality measurement and a good-quality stack object (bit 4 and 16), while rejecting those flagged as suspect or poor-quality stack measurements (bit 64 and 128). We additionally required the PSF-weighted fraction of unmasked pixels (\textit{QfPerfect}) to exceed 0.85 in each band. To exclude extended sources we only retained those that satisfied: $i_{\mathrm{PSF}} - i_{\mathrm{Kron}} < 0.05$.

Finally, we matched the UKIDSS, 2MASS, and Pan-STARRS catalogues within a tolerance of 1" to the IRAC catalogue. When a match was found for the UKIDSS or 2MASS $J$-, $H$- or $K_S$-bands, the corresponding photometry was added to the IRAC source. If both catalogues provided measurements in the same band, the value with the lower uncertainty was retained. Likewise, when a match with the Pan-STARRS catalogue was identified, the $g$, $r$, $i$, $z$ or $y$ band data were added to the corresponding IRAC source. Sources from the UKIDSS, 2MASS and Pan-STARRS catalogues without IRAC counterparts were disregarded.

When constructing the IRAC catalogue, multiple detections within the matching tolerance were rare. Among sources with detections in all four IRAC bands, only 15 had more than one candidate counterpart within the 1" tolerance and therefore required a minimization procedure to select the best match. For all other matches, only a single detection fell within the tolerance. The same was true when matching the IRAC catalogue to UKIDSS and 2MASS, where every match within 1" was unambiguous (i.e. no source had more than one candidate counterpart). For Pan-STARRS, only 2 sources had more than one detection within the tolerance, in which case the detection closest to the IRAC coordinates was adopted.

\subsection{Completeness}

\begin{figure}
    \includegraphics[width=\columnwidth]{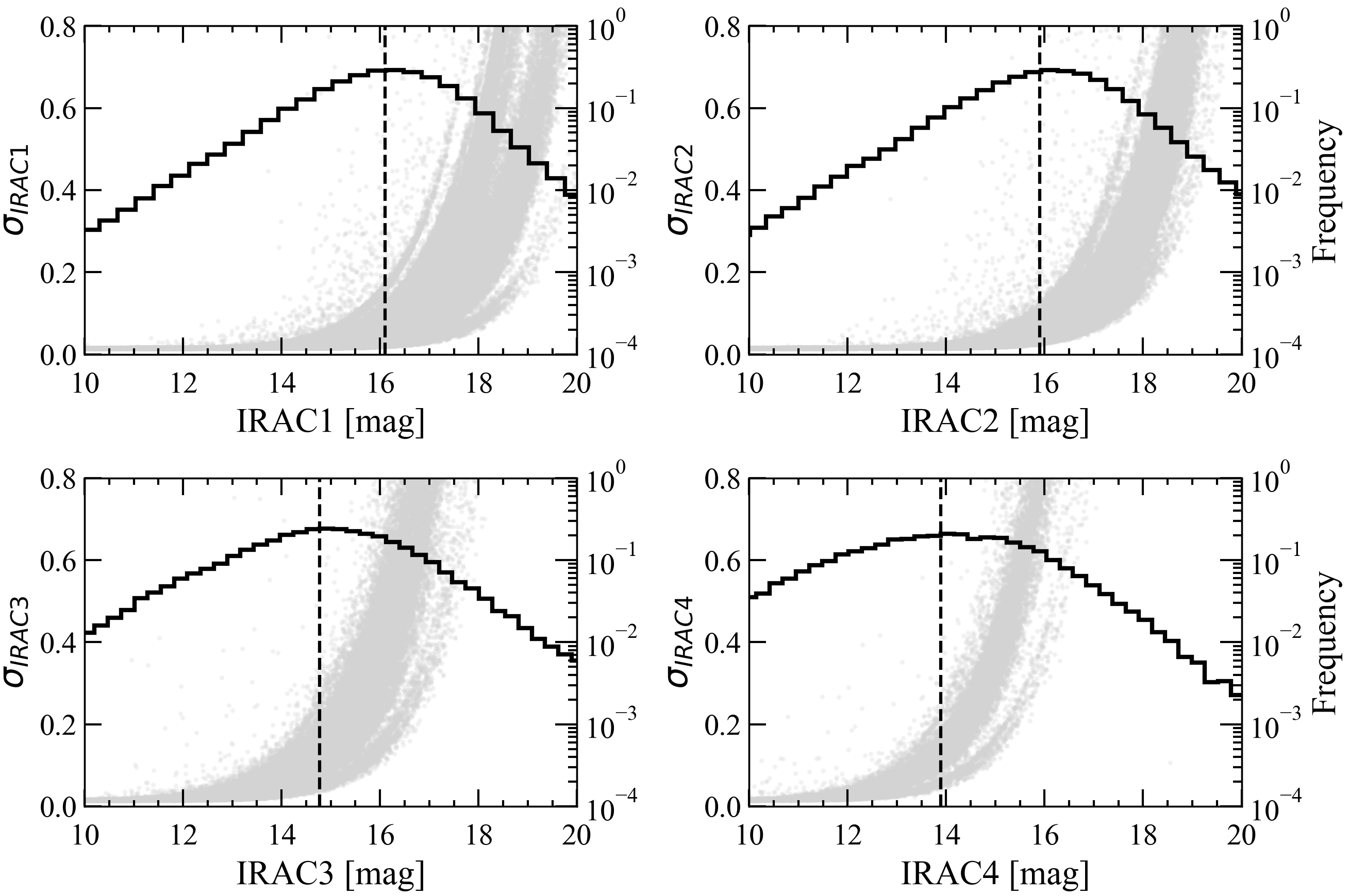}
    \caption{Photometric uncertainties (grey dots) and source density (black line) are shown as functions of magnitude for the IRAC1 (upper left), IRAC2 (upper right), IRAC3 (bottom left), and IRAC4 (bottom right), along with the completeness magnitude (dashed black line) for each filter.}
    \label{fig:IRAC_completness}
\end{figure}

\begin{figure}
    \includegraphics[width=\columnwidth]{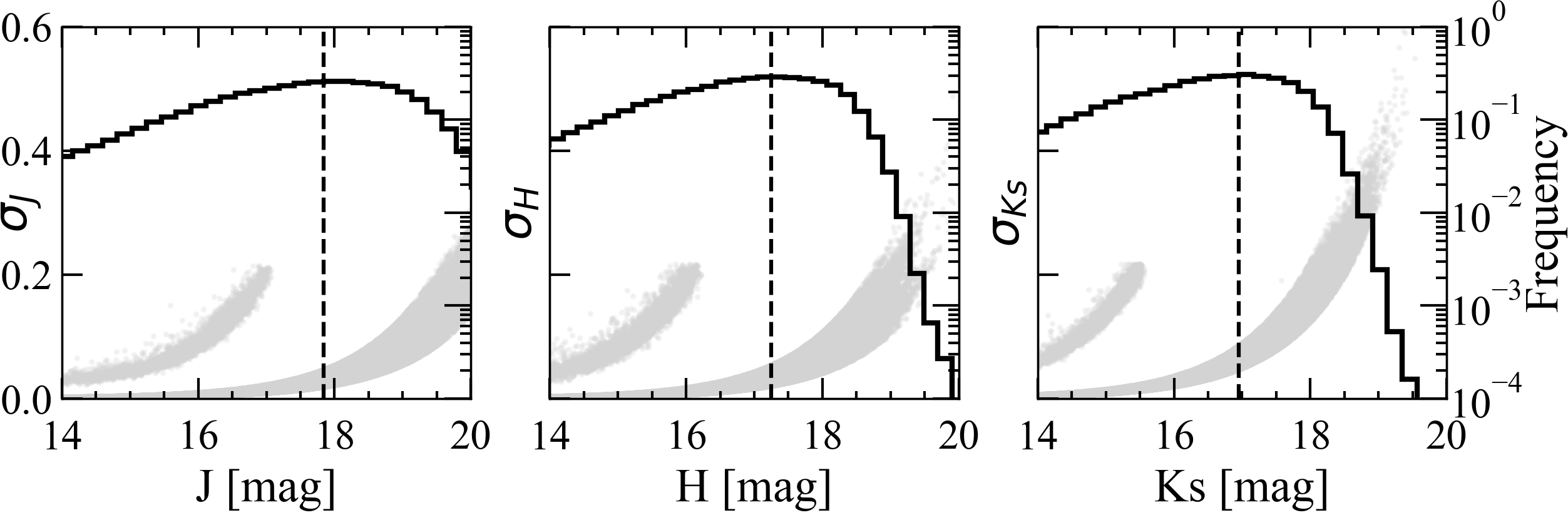}
    \caption{Photometric uncertainties (grey dots) and source density (black line) are shown as functions of magnitude for the J (left), H (middle), and K$_S$-bands (right), along with the completeness magnitude (dashed black line) for each filter.}
    \label{fig:JHKs_completness}
\end{figure}

We now have a catalogue spanning from mid-infrared to optical wavelengths covering the Rosette region. Our IRAC photometry is shown in Fig.~\ref{fig:IRAC_completness} as a function of photometric uncertainty, together with the frequency of objects in magnitude bins. The NIR photometry is presented in the same way in Fig.~\ref{fig:JHKs_completness}. In the NIR data, two distinct groups are visible, one originating from the 2MASS catalogue and the other from UKIDSS, with the latter being deeper. The peak in the number of sources is indicated by a black dashed line, representing the completeness magnitude. The completeness magnitudes for all photometric bands in our final catalogue are listed in Table~\ref{Photometric Completeness}. These values were determined using a bin size equal to one hundredth of the difference between the highest and lowest magnitudes in each photometric band.

\begin{table}
    \caption[Photometric completeness]{Photometric completeness for all filters.}
    \centering 
    \begin{tabular}{l c c}
        \toprule[1pt]
        Filter & $\lambda_{\text{eff}}$ [µm] &Completeness [mag] \\
        \midrule[1pt]
        IRAC4  & 8.0 & 13.9 \\
        IRAC3  & 5.8 & 14.8 \\
        IRAC2  & 4.5 & 15.9 \\
        IRAC1  & 3.6 & 16.1 \\
        $K_S$  & 2.2 & 17.0 \\
        $H$    & 1.7 & 17.2 \\
        $J$    & 1.2 & 17.8 \\
        $y$    & 1.0 & 19.6 \\
        $z$    & 0.9 & 19.8 \\
        $i$    & 0.8 & 20.3 \\
        $r$    & 0.6 & 21.5 \\
        $g$    & 0.5 & 21.7 \\
        \bottomrule[1pt]
        \end{tabular}

    \vspace{2mm}
    
    {\footnotesize 
    \textbf{Notes.} The $g$, $r$, $i$, $z$, $y$ magnitudes are in the AB magnitude system.}
    \label{Photometric Completeness}
\end{table}

Based on these figures, a photometric cutoff was applied in all bands. For IRAC1 and IRAC2, measurements with uncertainties above 0.2 magnitudes were disregarded, while for IRAC3 and IRAC4, photometric data with uncertainties above 0.4 magnitudes were excluded. These thresholds were chosen since they roughly correspond to the photometric uncertainties near the completeness limit. For the remaining bands, namely $g$, $r$, $i$, $z$, $y$, $J$, $H$ and $K_S$, measurements with uncertainties above 0.2 magnitudes were disregarded. Note that we are only removing data for a band where the uncertainty exceeds the chosen threshold -- if the source has valid photometry in other bands, it will remain in our catalogue. We note that the uncertainty cutoff in the optical/near-infrared bands is chosen to be consistent with IRAC1 and IRAC2. In practice, the exact value of this cutoff has little impact on the further results; the depth of the catalogue is determined by the IRAC images.

\begin{figure}
    \includegraphics[width=\columnwidth]{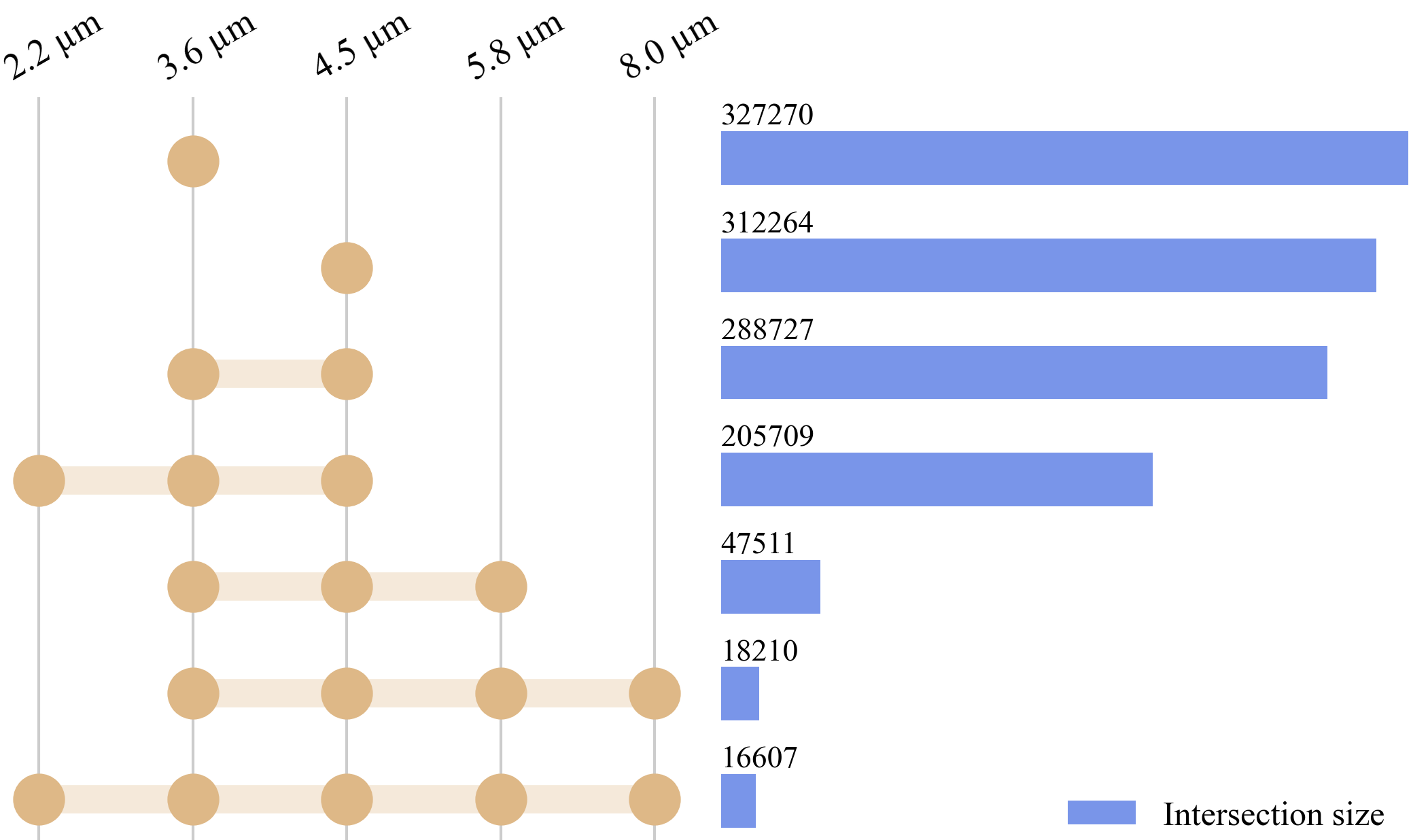}
    \caption{Diagram showing the number of detected sources after the magnitude uncertainty cutoff for various combinations of IRAC filters and the 2.2 µm $K_S$ filter.}
    \label{fig:NumberSources_cutoff}
\end{figure}

In Fig.~\ref{fig:NumberSources_cutoff} we show the resulting numbers of sources with detections for various combinations of IRAC filters and the 2.2$\,\mu  m$ $K_S$ band. The figure demonstrates that our catalogue contains 205,709 sources with valid photometry in $K_S$, IRAC1 and IRAC2, however, only 18,210 sources have data in all four IRAC channels. In the following, we only make use of this smaller sample with a complete set of IRAC photometry.

\section{Selection of young (sub)stellar objects}
\label{sec:selection}
\subsection{Colour-colour selection}

In this section, we describe a set of colour–colour criteria for the selection of YSOs, applied only to sources with IRAC photometry in all four bands. In total 18,210 sources were analysed, with uncertainties below 0.2 magnitudes for IRAC1 and IRAC2 and uncertainties below 0.4 magnitudes for IRAC3 and IRAC4.

Our selection is based on the 'Phase 1' classification scheme in Appendix A of \citet{Gutermuth}. The only difference is that we do not apply the colour–magnitude criteria shown in Fig. 14 in \citet{Gutermuth}. This cutoff is designed to remove faint extragalactic sources. However, in our deep fields near the Galactic plane, applying this cutoff would remove many genuine point sources that could be young stars. We note that this selection scheme explicitly includes error terms to account for variable photometry uncertainties.

\begin{figure}
    \includegraphics[width=0.98\columnwidth]{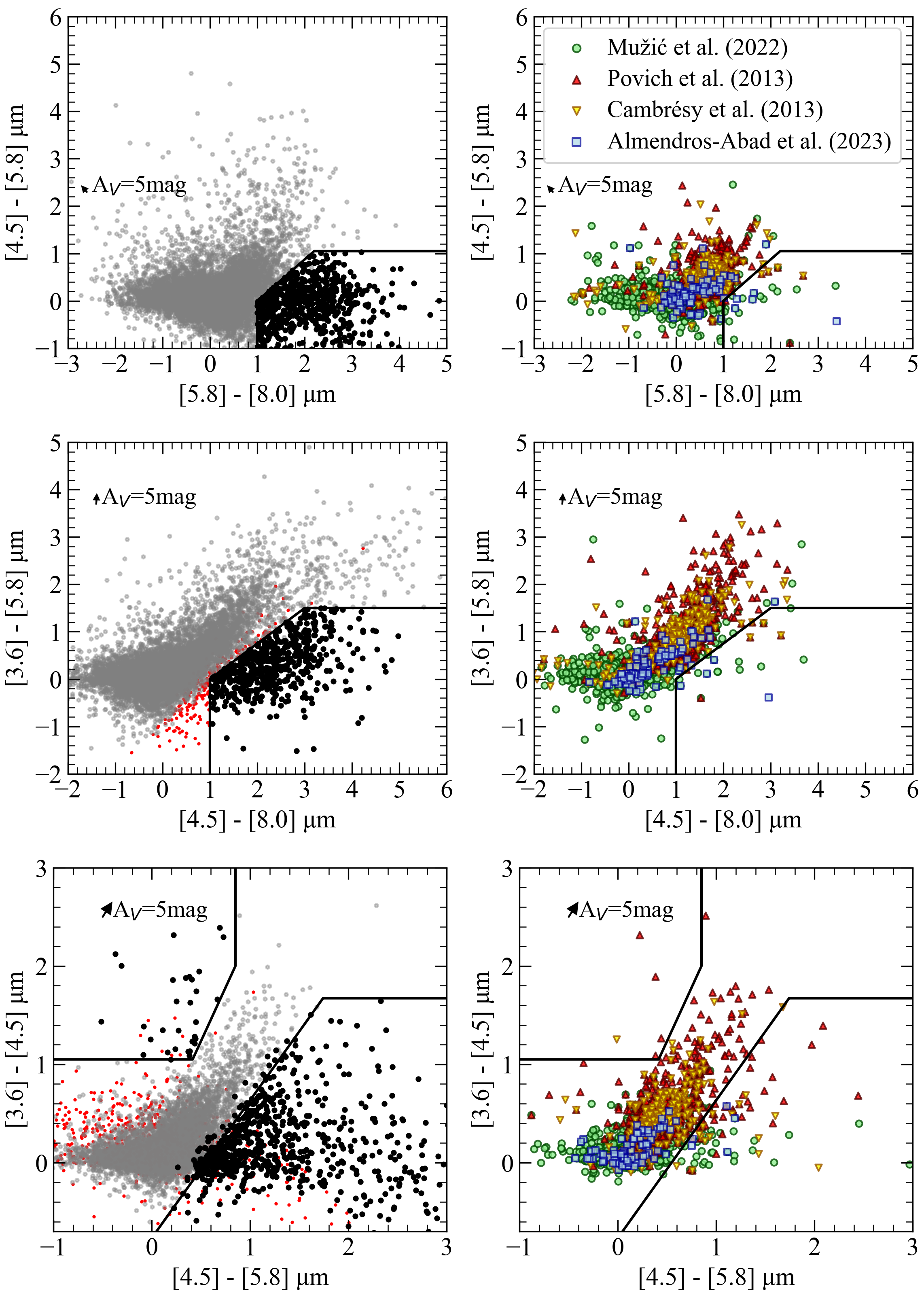}
    \caption{Colour–colour diagrams and the cutoffs as defined by the Phase 1 classification scheme in Appendix A of \citet{Gutermuth}. The three colour-colour diagrams illustrate the criteria used to exclude contaminant sources. On the left panels, our dataset (grey circles) is shown with selected sources represented by black circles. These highlighted sources are shown as red dots thereafter and disregarded from further classification. On the right, the same diagram shows sources from \citet{Cambresy} (yellow downward triangles), \citet{Povich} (red upward triangles), \citet{Muzic_2022} (green circles), and \citet{Almendros2023} (blue squares). An $A_V=5$ mag reddening vector is represented.}
    \label{fig:Type1}
\end{figure}

\begin{figure}
    \includegraphics[width=0.98\columnwidth]{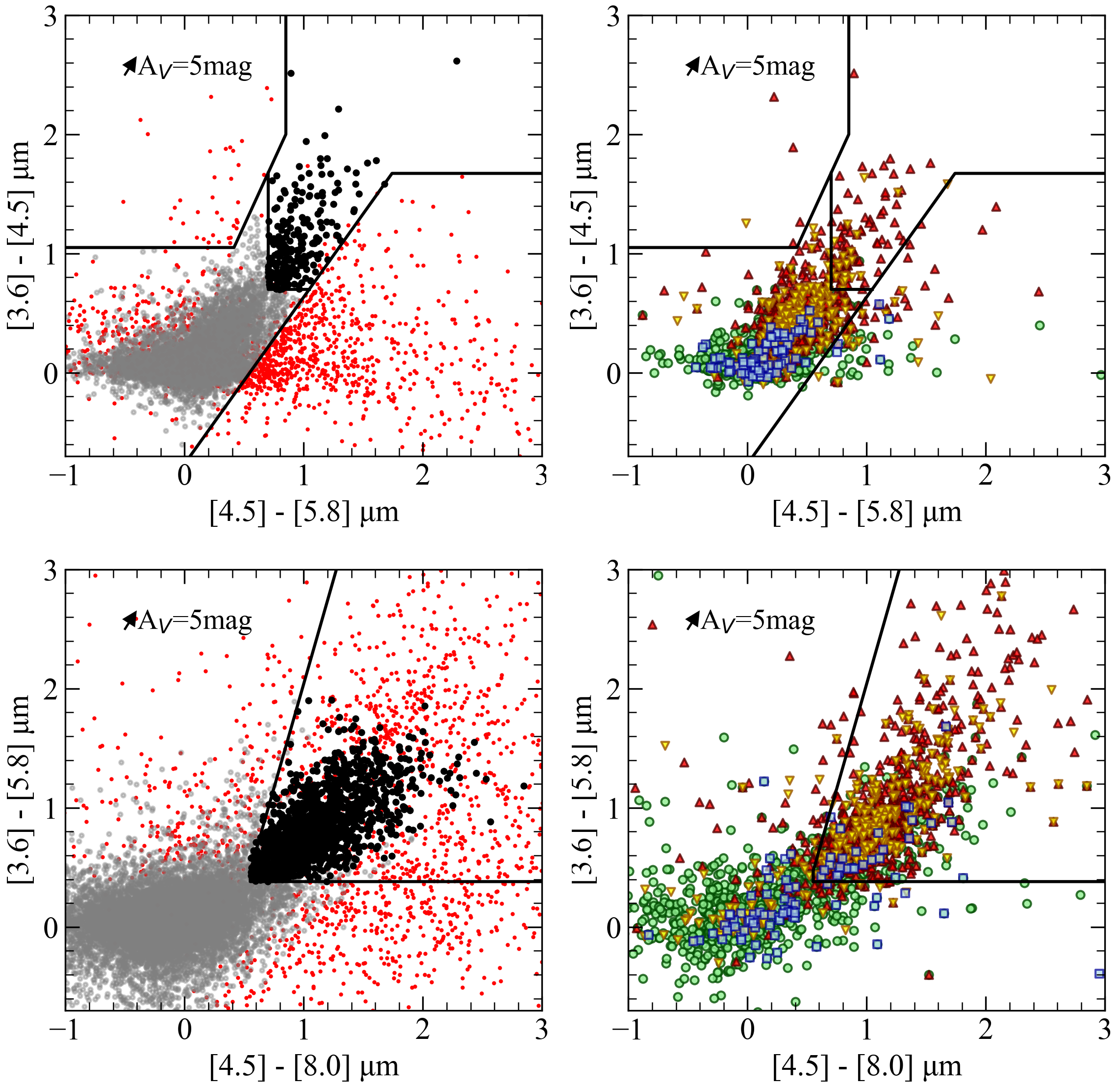}
    \caption{Colour–colour diagrams and the selections as defined by the Phase 1 classification scheme in Appendix A of \citet{Gutermuth}. The two colour-colour diagrams adopted to identify YSO candidates. On the left panels, our dataset (grey circles) is shown with selected sources represented by black circles. In the first row, the red dots represent previously excluded contaminants. In the second row, the red dots also represent sources already classified as Class I, which are excluded from further classification. On the right, the same diagram shows sources from \citet{Cambresy} (yellow downward triangles), \citet{Povich} (red upward triangles), \citet{Muzic_2022} (green circles), and \citet{Almendros2023} (blue squares). An $A_V=5$ mag reddening vector is represented.}
    \label{fig:Type2}
\end{figure}

In Fig.~\ref{fig:Type1} and Fig.~\ref{fig:Type2} we show the five colour–colour criteria used for our selection. In these figures, we also compare our selection with relevant sources from previous studies of the Rosette Nebula, hereafter referred to as the literature sample, which match our photometric catalogue within a 1" tolerance (right column of Fig.~\ref{fig:Type1} and Fig.~\ref{fig:Type2}). This literature sample includes infrared excess YSO candidates identified by \citet{Cambresy} and \citet{Povich}, which use MIR photometry from WISE \citep[Wide-field Infrared Survey Explorer]{AllWISE} and Spitzer, respectively. It also includes probable members of the Rosette Nebula from \citet{Muzic_2022}, as well as bona-fide and candidate members of NGC~2244 from \citet{Almendros2023}.

The five colour-colour cutoffs are applied to the catalogue sequentially. The first two colour–colour cutoffs (upper two rows in Fig.~\ref{fig:Type1}) remove active star-forming galaxies, characterized by their strong polycyclic aromatic hydrocarbon (PAH) feature emission at mid-infrared wavelengths. The next cutoff (third row in Fig.~\ref{fig:Type1}) is designed to remove both unresolved shock-emission knots (top left) and faint field stars whose photometry is contaminated by structured PAH emission (bottom right). In the fourth colour–colour diagram (first row in Fig.~\ref{fig:Type2}), which is in fact the same as the third one, we specifically select Class I YSO candidates. Finally, the fifth colour–colour diagram (second row in Fig.~\ref{fig:Type2}) is used to select Class II YSO candidates. For further details on these criteria, we refer to \citet{Gutermuth}.

\begin{figure}
    \includegraphics[width=\columnwidth]{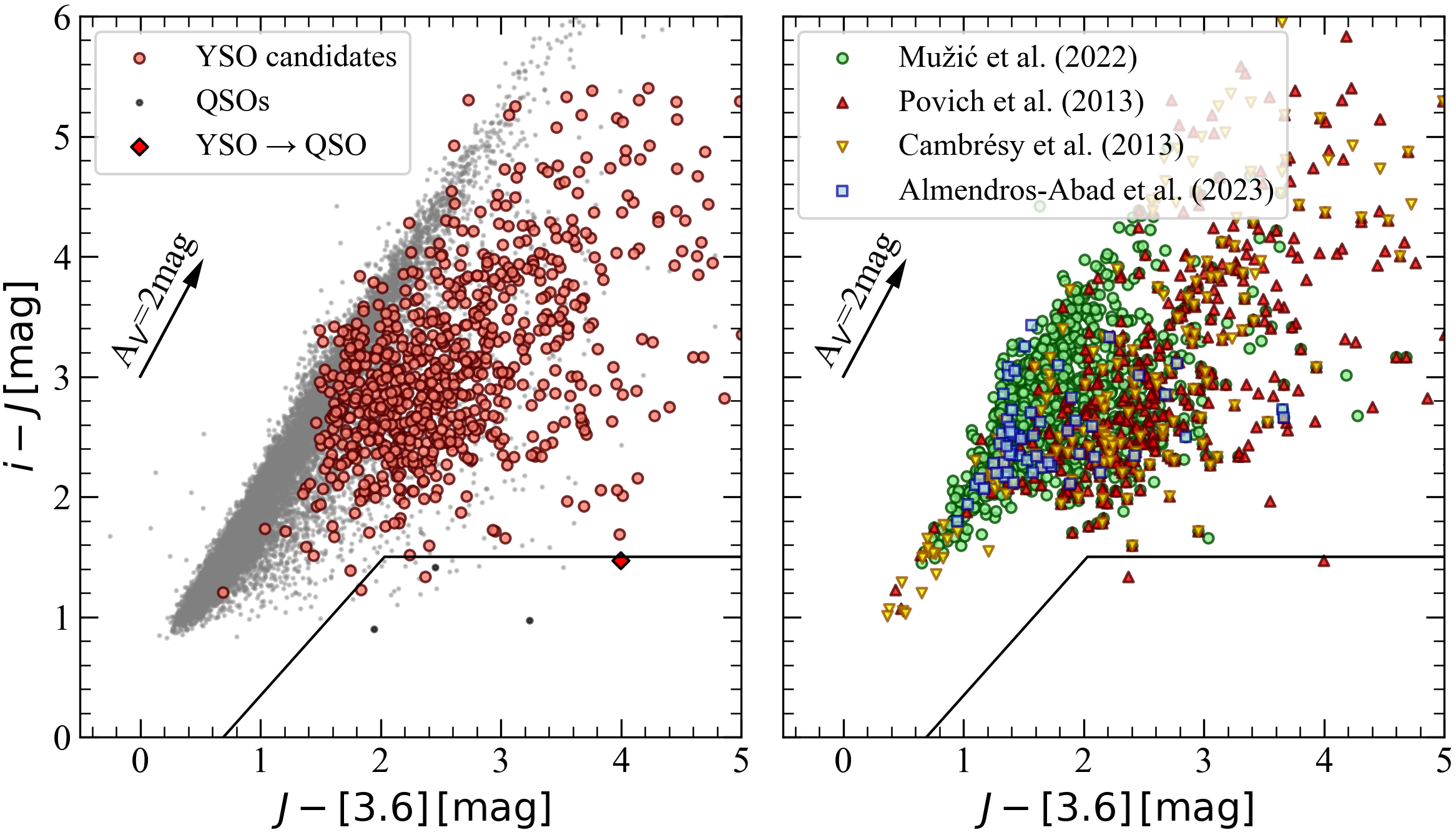}
    \caption{Colour–colour diagram and the selection represented by Eq.~\ref{equation_qso} (black line). On the left side, our dataset (grey dots) is displayed, with sources satisfying the equation marked as black circles. Our YSO candidates are shown as red circles, while those reclassified as QSOs are indicated by red diamonds with black outlines. On the right, the same diagram shows sources from \citet{Cambresy} (yellow downward triangles), \citet{Povich} (red upward triangles), \citet{Muzic_2022} (green circles), and \citet{Almendros2023} (blue squares). An $A_V = 2$ mag reddening vector is shown.}
    \label{fig:QSO_TypeI}
\end{figure}

Additionally, we verify YSO candidates using a criterion similar to the Quasi-Stellar Object (QSO) cutoff proposed by \citet{Bouy}. This cutoff is applied only to sources with photometric data in both the $J$ and $i$ bands with uncertainties below 0.2 magnitudes. Of the 1529 YSO candidates (1304 Class II and 225 Class I), 783 (767 Class II and 16 Class I), about half, meet this requirement. Equation~\ref{equation_qso} shows the criterion for a source to be considered a QSO, and these constraints are illustrated in Fig.~\ref{fig:QSO_TypeI}. This results in the reclassification of a single YSO candidate. Given that only about half of the catalogue have the required photometry to use this criterion, there could be a very small number of unrecognised QSOs in our sample.

\begin{equation}
\begin{split}
\hspace*{2.5cm} 
\mathrm{[3.6]} &> 13 \\ 
\mathrm{i} - \mathrm{J} &\leq 1.5 + \sigma[\mathrm{i} - \mathrm{J}] \\
\mathrm{i} - \mathrm{J} &< 1.12 \times (\mathrm{J} - \mathrm{[3.6]}) - 0.8
\end{split}
\label{equation_qso}
\end{equation}

The selection yielded a total of 1528 YSO candidates. The classification of the analysed sources is summarized in Table~\ref{tab:TypeI_results}.

\begin{table}
    \caption{Classification results for all analysed sources.}
    \centering
    \begin{tabular}{l c}
        \toprule[1pt]
        Classification & \#Sources \\
        \midrule[1pt]
        YSO Class II & 1303 \\
        YSO Class I & 225 \\
        Unclassified & 15085 \\
        PAH contamination & 782 \\
        PAH Galaxy & 782 \\
        Shock Emission & 33 \\
        QSO & 1 \\
        \bottomrule[1pt]
    \end{tabular}
    \label{tab:TypeI_results}
\end{table}

\subsection{Review of YSO candidates}

Our sample consists of 1303 Class II and 225 Class I YSOs. As designed, the used colour-colour criteria will preferentially select sources with disks or embedded sources, and will neglect diskless young stars. 

On the right side of Fig.~\ref{fig:Type1} and Fig.~\ref{fig:Type2} we show how the literature sample relates to the adopted selection criteria. The overwhelming majority of previously identified YSOs survive the first three cuts, as expected. Moreover, the majority of the sources from \citet{Cambresy} and \citet{Povich} lie within the regions corresponding to our Class I and Class II selections, consistent with their expected nature. Furthermore, many of the sources from \citet{Muzic_2022} and \citet{Almendros2023}, which are expected to consist predominantly of Class III YSOs, occupy the expected regions of colour-colour space for diskless stellar objects.

\begin{table}
    \caption[comparison results]{Details of the recovery rate of YSO candidates.}
    { 
    \centering
    \begin{tabular}{l c c}
        \toprule[1pt]
         & \#Sources & \#YSO candidates \\
        \midrule[1pt]
        \citet{Cambresy} & 306 & 192 (63\%)  \\
        \citet{Povich} & 707 & 574 (81\%)  \\
        \citet{Muzic_2022} & 1576 & 406 (26\%)  \\
        \citet{Almendros2023} & 94 & 22 (23\%)  \\
        \bottomrule[1pt]
    \end{tabular}
    }
    
    \vspace{2mm}
    
    {\footnotesize 
    \textbf{Notes.} The first column lists how many sources from each catalogue are included in the 18,210 objects analysed. The second column gives the number of sources classified as YSO Class I or Class II, together with their corresponding recovery fractions.}
    \label{tab:TI_results}
\end{table}

In Table~\ref{tab:TI_results} we report the recovery rates for the four literature sample catalogues. For each sample, we list the number of objects in our photometric catalogue, and the number that we selected as YSOs. In total, about half of our YSO sample (804 sources) coincide with objects from the literature sample. The recovery rate is above 60\% for previous infrared surveys, and below 30\% for surveys based on optical/near-infrared data. The highest recovery rate of 81\% is found for the study by \citet{Povich}, which also used Spitzer data, albeit with a different methodology, providing re-assurance in our YSO selection. Comparing with \citet{Cambresy}, the recovery rate is 63\%. This drop-off can be explained by the fact that these authors use WISE data. On one hand, the poor spatial resolution of WISE may lead to spurious sources and blended photometry. On the other hand, the YSO selection in that paper extends to wavelengths of 12 and 22$\,\mu m$.

\begin{figure*}
    \centering
    \includegraphics[width=\textwidth]{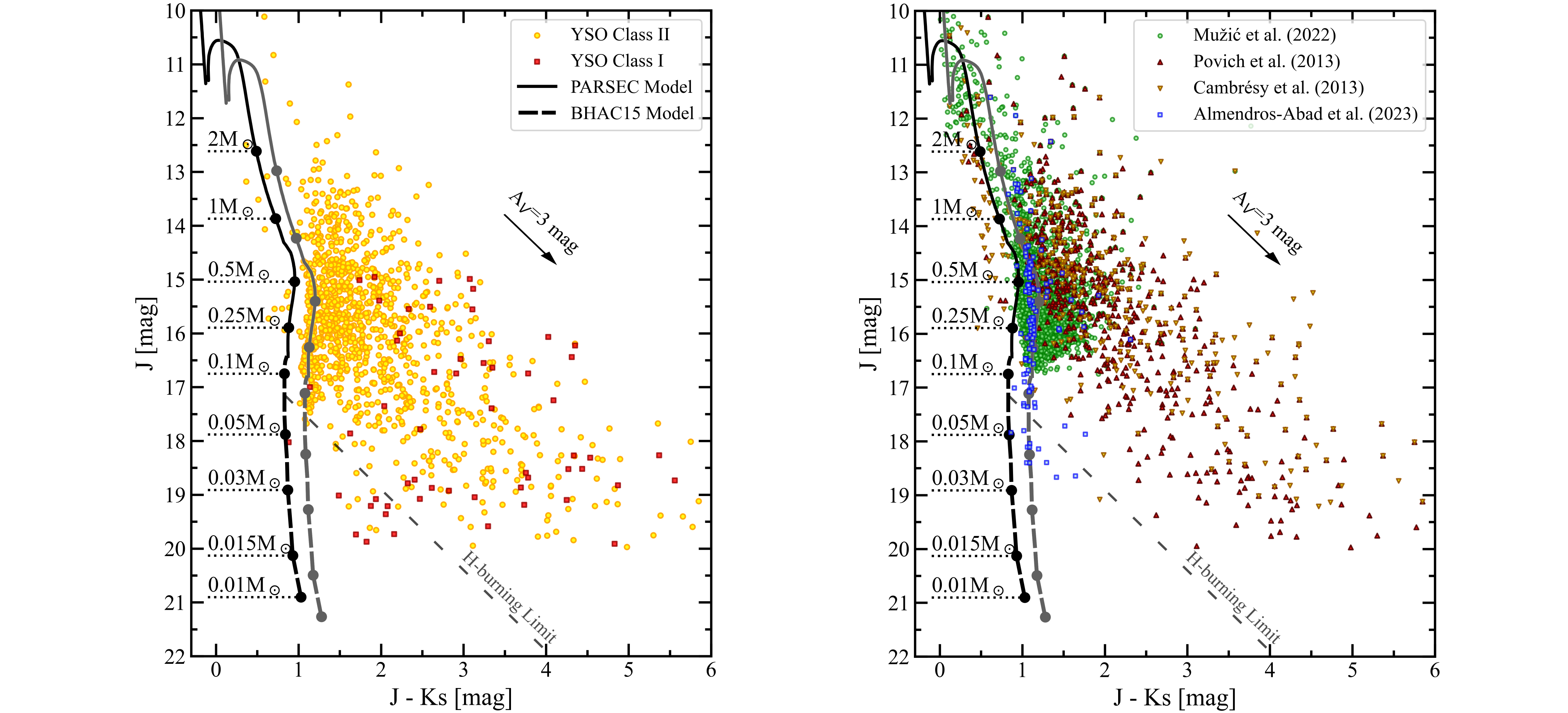}
    \caption{Colour-magnitude diagram of the YSO candidates (left panel) and literature sample (right panel). In the left panel, we show the selected YSO candidates, with Class I YSOs represented as red squares and Class II YSOs as orange circles. In the right panel we compare our results with infrared-excess YSO candidates from \citet{Cambresy} (yellow downward triangles) and \citet{Povich} (red upward triangles), probable Rosette members from \citet{Muzic_2022} (green circles), and bona-fide and candidate NGC~2244 members from \citet{Almendros2023} (blue squares). The 2 Myr PARSEC isochrone \citep{Bressan,Prisinzano} is plotted as a solid black line, and the 2 Myr BHAC15 model \citep{BHAC15} as a dashed black line, both at a distance of $1500\ \mathrm{pc}$. Additionally, the same isochrones are shown in grey for an extinction of $A_V=1.5$ (the average extinction of NGC~2244; \citealp{Muzic2019}). The thin dashed line indicates the hydrogen-burning limit, objects below this line are classified as substellar. A reddening vector corresponding to $A_V=3\ \mathrm{mag}$ is also shown.}
    \label{fig:CMD_result}
\end{figure*}

The YSO candidates are displayed in a colour–magnitude diagram in Fig.~\ref{fig:CMD_result}, in comparison with a theoretical isochrone. To re-iterate, only sources with $J$- and $K_S$-band photometry with uncertainties of 0.2 mag or lower are displayed, which excludes 383 out of the total 1528 YSO candidates. The left panel of Fig.~\ref{fig:CMD_result} shows our YSO candidates. Most YSOs span a mass range between 2$\,M_{\odot}$ and the hydrogen-burning limit ($\sim$0.08$\,M_{\odot}$). A noticeable trend is that Class I YSO candidates exhibit stronger reddening, consistent with these objects being deeply embedded.

The right panel of Fig.~\ref{fig:CMD_result} presents objects from the literature sample. Sources from \citet{Cambresy} and \citet{Povich} exhibit relatively high extinction, with distributions that resemble our YSO sample (left panel of Fig.~\ref{fig:CMD_result}). However, our sample is substantially deeper, by at least 1\,mag in J-band, at low extinction, and several mag at higher values of $A_V$. Therefore our sample includes a substantial number of objects that may be very low mass stars and brown dwarfs based on their position in the colour-magnitude diagram. In fact, our sample includes 22 very low mass sources confirmed spectroscopically by \citet{Almendros2023}, two of which are brown dwarfs according to their spectral type. In contrast, sources from \citet{Muzic_2022} and \citet{Almendros2023} present overall lower extinction, lying closer to the main sequence compared to the other YSO candidates, reflecting their more evolved nature.

\begin{figure}
    \includegraphics[width=\columnwidth]{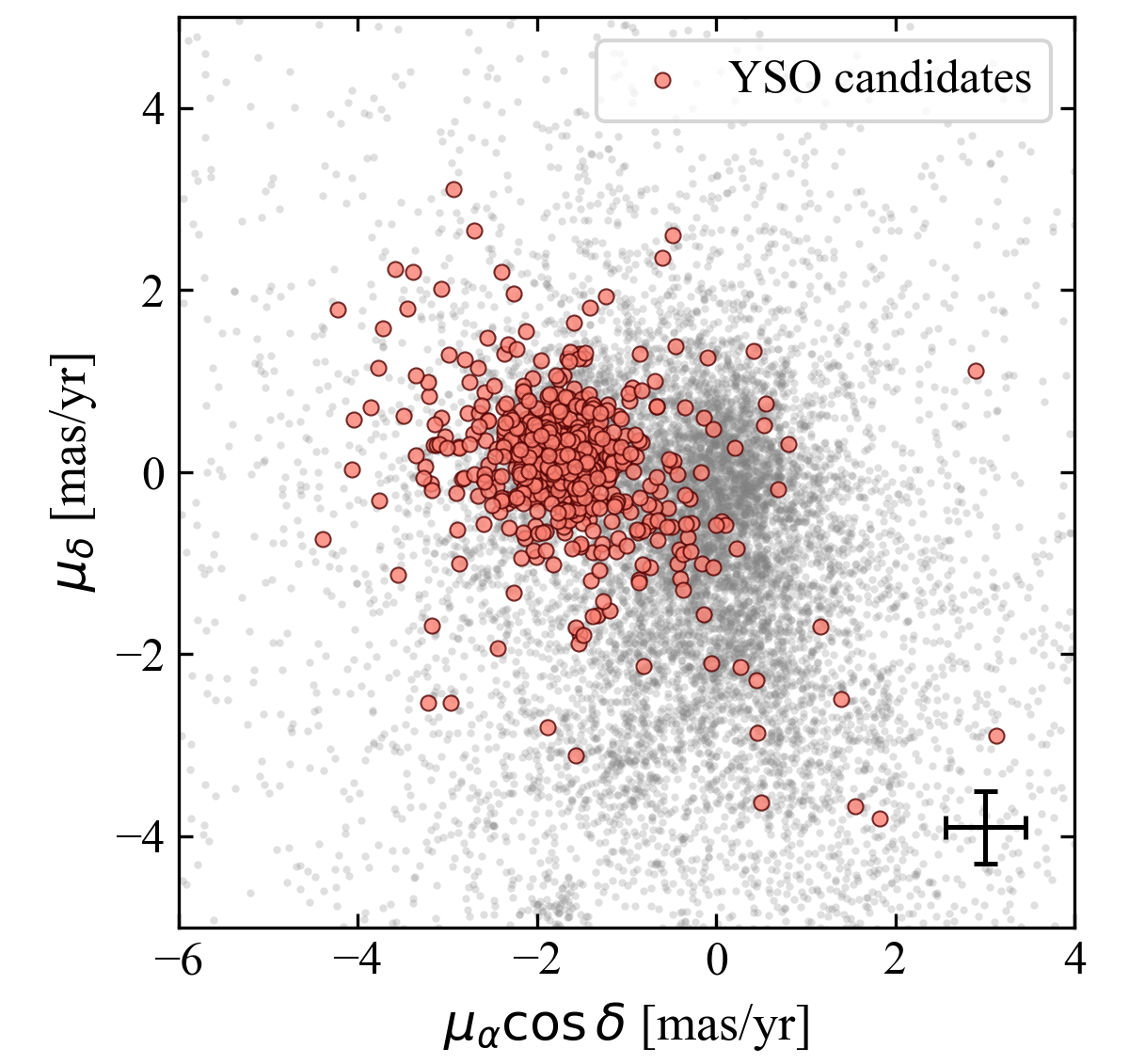}
    \caption{ Gaia proper motions for sources detected in all four IRAC bands. YSO candidates are shown in red, the remaining sources are shown as grey points. The mean proper motion uncertainty of the YSO sample is shown in the lower right corner.}
    \label{fig:GAIApm}
\end{figure}

As an additional check, we cross-matched the 18,210 analysed sources with the Gaia DR3 catalogue \citep{GaiaDR3} using a 1" tolerance. Of these, 13,969 objects have a Gaia counterpart with proper motion measurements and RUWE < 1.4, of which 614 are YSO candidates. The proper motions of these YSO candidates shown in Fig.~\ref{fig:GAIApm}, are strongly clustered around the values expected for young stars in this region, $\sim (\mu_{\alpha}\cos\delta = -2, \mu_{\delta} = 0) \text{ [mas/yr]}$ \citep{Muzic_2022}. This provides re-assurance that the selection is indeed identifying overwhelmingly young stellar members of the nebula, rather than foreground or background contaminants. Confirmation of individual objects, however, will require follow-up spectroscopy.

In summary, our survey yields 1528 YSO candidates, including 669 new objects not reported in the literature sample (see above) or in other relevant surveys \citep{Bell13, Broos, Meng_2017}. Although our survey does not extend across the entire Rosette Nebula, we cover the regions where YSOs are expected to be most abundant, associated with known stellar groups. Restricting the analysis to sources with photometry in all IRAC bands enabled the selection of a reliable sample of YSO candidates. In addition, this photometric coverage allowed us to distinguish between Class I and Class II YSOs.

\section{Spatial distribution of young stars}
\label{sec:Sec4}
\subsection{Overview}

\begin{figure*}
    \centering
    \includegraphics[width=\textwidth]{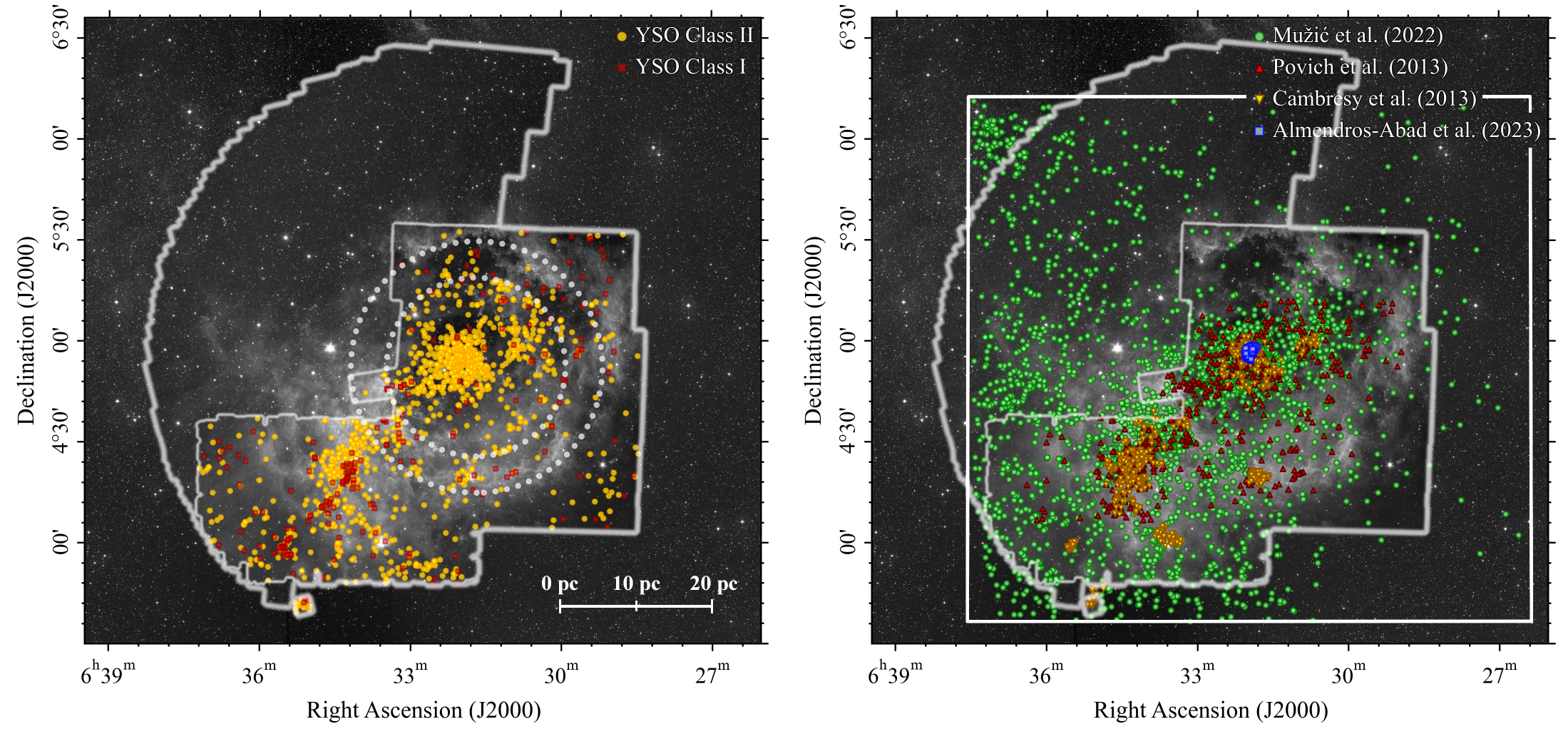}
    \caption{Spatial distribution of YSOs in the Rosette Nebula overplotted on a DSS background combined with part of the IRAC2 mosaic. The area covered by IRAC1 and IRAC2 is outlined with a thick grey line, while the region covered from IRAC1 to IRAC4 is outlined with a thin grey line. Two dotted circles are also shown, the smaller one marks the cavity region, while the annular area between the two circles denotes the Interaction ring. The left panel shows Class I (red squares) and Class II (orange circles) YSO candidates. The right panel presents YSOs from the literature sample, \citet{Cambresy} (yellow downward triangles), \citet{Povich} (red upward triangles), \citet{Muzic_2022} (green circles), and \citet{Almendros2023} (blue squares), along with a white rectangle indicating the spatial coverage of \citet{Muzic_2022}.}
    \label{fig:Result_distribution}
\end{figure*}

The spatial distribution of our YSO candidates and the literature sample is displayed in Fig.~\ref{fig:Result_distribution}. Sources from the literature sample are shown in full, whether or not they match objects in our photometric catalogue. The area covered by IRAC1 and IRAC2 is marked with a grey line, while the coverage in IRAC1 to IRAC4 is outlined with a thin grey line, within which all YSO candidates fall (see left panel of Fig.~\ref{fig:Result_distribution}). The right panel of Fig.~\ref{fig:Result_distribution} shows the locations of the sources from the literature sample, along with a white rectangular outline indicating the survey footprint of \citet{Muzic_2022}.

Although young stars are found throughout the surveyed region, their densities appear to vary widely. The overall distribution of our YSO candidates indicates clustering. This suggests a low level of background and foreground contamination in our sample. The majority of YSOs are concentrated in the central cavity of the Rosette Nebula, where NGC~2244 is located, and in the central core of the Rosette Molecular Cloud, situated $\sim$16-23~pc to the south-east of the centre of NGC~2244. We further evaluate the clustering of YSOs in Section \ref{sec:density}.

The spatial distribution of different YSO classes provides clues into the star formation history of a region. Since Class I YSOs represent an earlier evolutionary stage than Class II objects, regions with a higher fraction of Class I sources are expected to be younger. From the left panel of Fig.~\ref{fig:Result_distribution}, it is strongly suggestive that the spatial distributions of Class I and Class II YSOs are significantly different. In fact, the median Class I source lies nearly twice as far from the cluster centre as the median Class II source, at 21 and 12~pc, respectively. For example, the central cavity (defined here as a circular region centred at $\alpha = 06{:}31{:}41.0$, $\delta = +04{:}52{:}00$, with an area of 2100~arcmin$^2$; smaller dotted circle in Fig.~\ref{fig:Result_distribution}) is dominated almost entirely by Class II YSOs, whereas several stellar groups in the south-east show a high fraction of Class I objects, indicating they are younger. Based on the visual examination of this figure, the youngest regions are primarily located away from the central cavity.

In the central core of the Rosette Molecular Cloud, some segregation between Class I and Class II YSOs is apparent. In this region, Class I YSOs tend to cluster toward the western part of the complex. This is explored in more detail in subsequent sections (see Sec. \ref{sec:stellargroups}).

Another noticeable feature is the presence of YSOs, predominantly of Class I, distributed around the central cavity in an annular or ring-like configuration. This region is indicated in Fig.~\ref{fig:Result_distribution} by a white dotted outline, where the Class I ratio is $\sim$30\%. This Interaction ring is the area directly beyond the cavity, extending out to an additional area of 2100~arcmin$^2$ -- equivalent to an annulus spanning $\sim$11-16~pc. This structure largely coincides with the region where the expanding HII front, driven by the OB association in NGC~2244, interacts with the surrounding molecular cloud.

The presence of YSOs beyond the interface region suggests that star formation in the Rosette Nebula can also proceed independently of the influence of NGC~2244 (more on this in Sec.~\ref{sec:chronology}). Moreover, outside these regions the YSO distribution appears highly inhomogeneous and exhibits filamentary structures, with loosely connected chains of YSOs linking denser groups (see Sec.~\ref{sec:density} for more details). This is consistent with the findings of \citet{Schneider}, who showed that star formation in the Rosette Nebula predominantly occurs along filaments.

\subsection{YSO density distribution}
\label{sec:density}

The spatial distribution of YSOs across the Rosette Nebula bears the fingerprints of its formation history, with visual inspection alone revealing apparent signatures of clustering and an underlying filament structure. A Clark-Evans nearest-neighbour test shows the observed YSO separations are only $\sim$70\% of the random-distribution expectation, with a deviation of more than 20$\sigma$, clearly indicating a clustered distribution. Additionally, a two-dimensional Kolmogorov–Smirnov test reveals that Class I and Class II YSOs cluster in different regions (D = 0.29, $p < 10^{-4}$).

\begin{figure}
    \centering
    \includegraphics[width=\columnwidth]{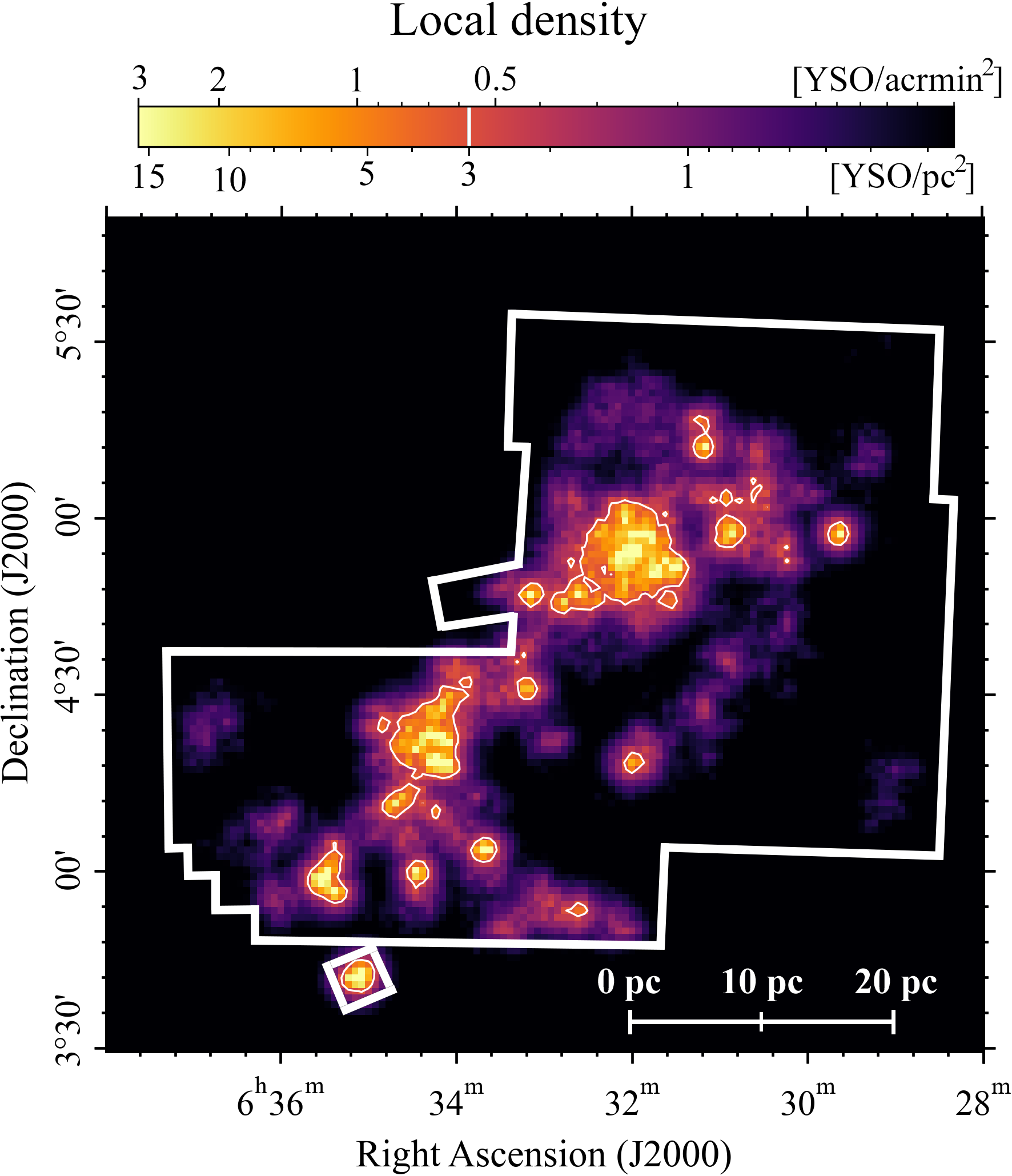}
    \caption{Map of the local surface density of YSOs in the Rosette Nebula, with a pixel grid on a $0.5\times0.5$~pc scale. A thin white contour outlines regions with a local density of at least 3~YSO~pc$^{-2}$. The thick white line traces the smoothed boundary of the survey footprint. A parsec scale for a distance of 1500~pc is shown in the bottom right.}
    \label{fig:isodensity}
\end{figure}

A map of the local surface density of YSOs across the Rosette Nebula is shown in Fig.~\ref{fig:isodensity}. We used the $N$th-nearest-neighbour estimator \citep{Casertano1985}, in which the surface density at any point is $\Sigma = (N-1)/(\pi\, d_N^2)$, where $d_N$ is the projected distance to the $N$th-nearest YSO. We adopted $N = 7$, following \citet{Bressert2010}. The density field was evaluated on a regular grid of $0.5 \times 0.5$~pc cells, and overdensities of at least 3~YSO~pc$^{-2}$ are outlined by a white contour. Here we begin to see the presence of several possible stellar clusters and groups, but perhaps more importantly, we can also discern what appears to be the underlying filamentary structure along which star formation occurred. A structure appears to connect NGC~2244 to the central core of the Rosette Nebula, oriented roughly north-west to south-east. Further from the centre, this filament seems to branch out and become more diffuse.

\begin{figure}
    \centering
    \includegraphics[width=\columnwidth]{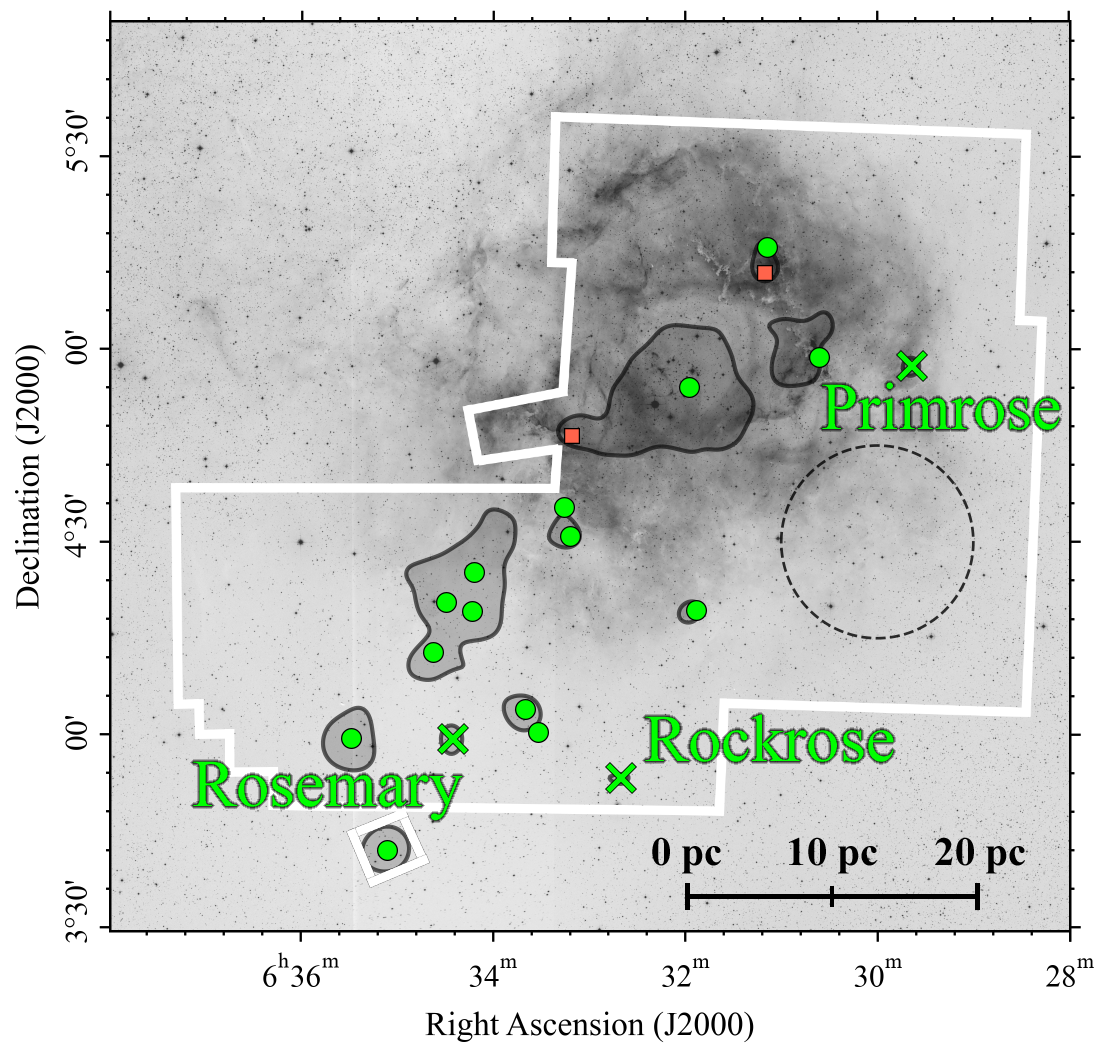}
    \caption{Kernel density estimation overdensity contours of the YSO candidates overplotted on a DSS background image. The control field (dark dashed circle) is defined as a circle with a radius of $15'$, centred at $\alpha = 06{:}30{:}00$ and $\delta = +04{:}30{:}00$. Regions with densities 10 times higher than the control field are highlighted by a black contour. Additionally, the positions of previously known young stellar groups covered in all IRAC bands are indicated by green circles, while newly identified young stellar groups are labelled accordingly and marked with small green crosses. Other notable overdense structures are shown as pale red squares.}
    \label{fig:KDE_rosette}
\end{figure}

To identify the locations of young stellar groups across the Rosette Nebula, we performed a kernel density estimation using the positions of the YSO candidates (see Fig.~\ref{fig:KDE_rosette}). The density was estimated on a $500\times500$ grid using a Gaussian kernel, with the bandwidth set to one quarter of Scott's rule ($n^{-1/6}$) applied independently to the RA and Dec standard deviations. We defined a control field, located away from regions with high YSO concentrations, to estimate the background YSO density. This control field is indicated by a black dashed circle with a radius of $15'$, centred at $\alpha = 06{:}30{:}00$ and $\delta = +04{:}30{:}00$. Regions with densities 10 times higher than the control field are then highlighted by black contours, a threshold chosen to suppress diffuse filamentary structure while preserving the significant, well-defined regions of overdensity. We note that choosing a different control field, or a mean average across the entire field, would produce the same result for the right threshold limit.

The majority of YSO overdensities trace the positions of previously known stellar groups \citep{PL7, Roman-Zuniga, Poulton}, marked with green circles. Notably, we recover all previously known young stellar groups, except forPL2 and PL3, which are not fully covered by IRAC4. We also seem to recover another group close to the position of REFL10, but it is unclear whether it actually corresponds to it, represents a subpart of it, or is a distinct, nearby stellar group. For simplicity, we adopt the label REFL10 throughout. In Fig.~\ref{fig:KDE_rosette} this is marked with a pale red square roughly north of NGC~2244.

A second pale red square, to the south-east of the centre of NGC~2244, marks one of the most identifiable features of the Rosette Nebula, a giant cold molecular pillar. At its tip we find a high concentration of YSOs at the interaction front with the expanding HII bubble. This is the same structure highlighted in Fig.~\ref{fig:NGC2244}, which, by visual inspection, appears to interact with the HII front in a manner similar to PL2 (Fig.~\ref{fig:AllG}\,\textit{b)}).

In addition, a few YSO overdensities appear at positions not associated with known stellar groups. Three of these, marked with small green crosses, are candidates for newly discovered young stellar groups. They were given the following thematic names, ordered by distance from NGC~2244: Primrose, Rockrose, and Rosemary. Primrose lies almost directly to the west of NGC~2244, Rockrose to the south, and Rosemary to the south of the central core of the Rosette Molecular Cloud, amidst other known stellar groups. The central coordinates, areas, and IRAC1 mosaic images of the regions occupied by these newly identified stellar groups are presented at the end of Appendix~\ref{clusters}.

Based on the work of \citet{Schneider}, the region occupied by Rosemary lies at a filamentary junction between the filament connecting the PL4–PL5–REFL8 complex, PL6, and REFL9, and another filament extending from PL3 and PouD. This is particularly interesting given that Rosemary appears as a chain of YSOs (see Fig.~\ref{fig:AllG} \textit{h)}).

\subsection{Minimum spanning tree clustering}
\label{sec:mst}

To test whether the groups identified in Sec.~\ref{sec:density} are statistically significant and recovered independently by clustering algorithms, we applied a Minimum Spanning Tree (MST) to the YSO candidates. The MST is the unique set of edges connecting all sources with no closed loops and the shortest possible total length, and its branch-length distribution provides a parameter-free way to separate clustered from distributed populations.

The critical branch length, separating intra-cluster branches from those bridging groups or connecting field stars, was found to be $L_{\rm crit} = 0.57$\,pc. Following \citet{Gutermuth}, this was taken as the intersection of straight-line fits to the steep (short, intra-cluster) and shallow (long, bridging) sections of the cumulative branch-length distribution. Clusters were then extracted by cutting every branch longer than $L_{\rm crit}$ and retaining the surviving connected components with at least 8 members.

\begin{figure}
    \centering
    \includegraphics[width=\columnwidth]{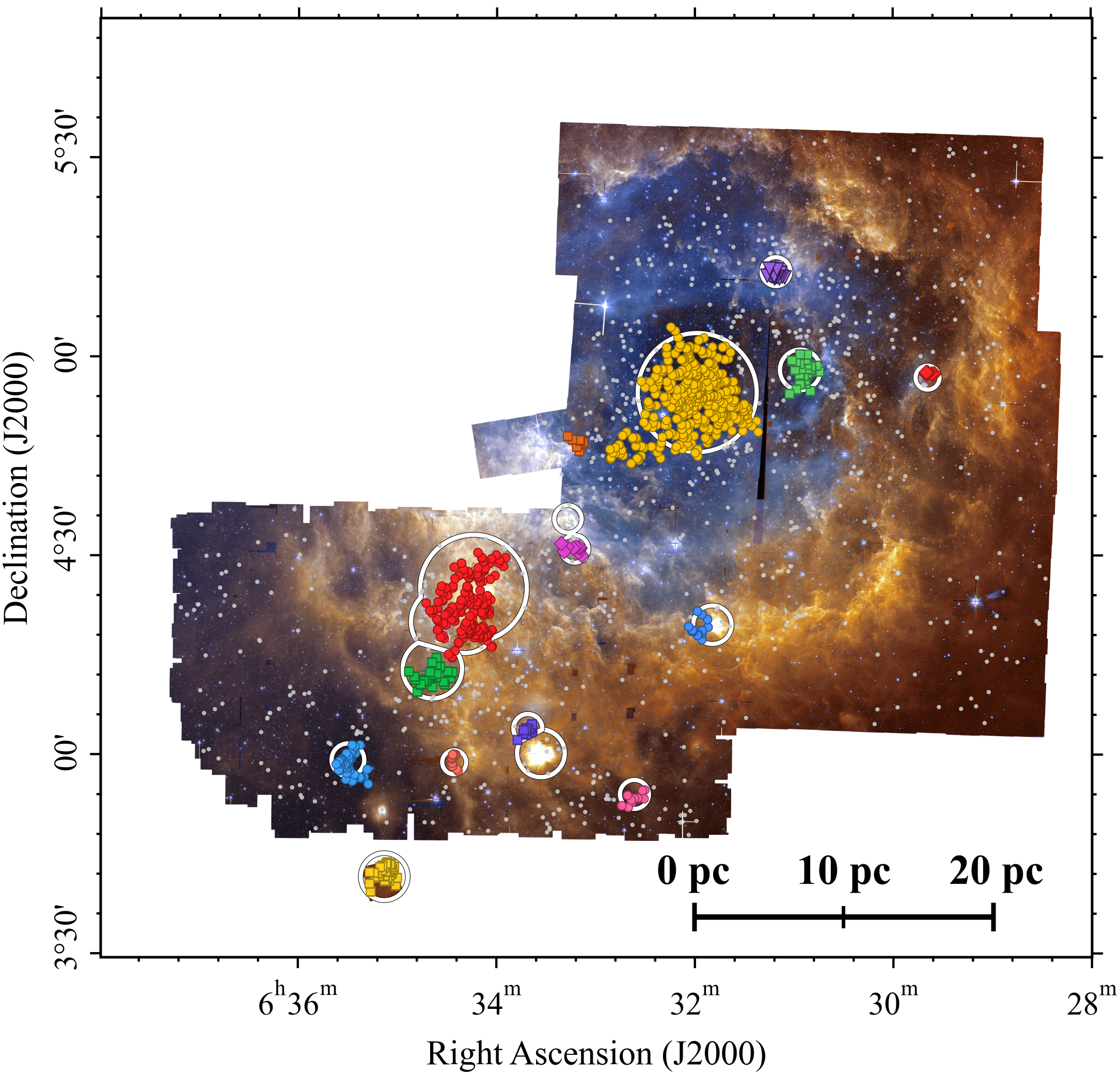}
    \caption{Spatial distribution of the YSO groups recovered by the minimum
    spanning tree analysis, overplotted on a composite image of the Rosette
    Nebula (Red: IRAC4, Green: IRAC3, Blue: IRAC2). Each significant group is shown in a
    distinct colour and/or marker; ungrouped YSO candidates are plotted as grey
    points. White outlines mark the regions discussed in Sec.~\ref{sec:stellargroups}.
    A parsec scale for a distance of 1500\,pc is shown in the bottom right.}
    \label{fig:MST}
\end{figure}

To confirm that each group is a real overdensity rather than a chance alignment of field stars, we estimated the field surface density as the median of the local $k$-nearest-neighbour density ($k=15$) over the survey footprint, giving $0.353$\,YSO\,pc$^{-2}$, and for each group computed the Poisson probability $P_{\rm bkg}$ of finding its observed number of members in its area by chance. Retaining only groups with $\log_{10} P_{\rm bkg} < -7.5$ leaves 14 significant groups. A small clump on the south-eastern edge of NGC~2244 was merged into it, as separating it would artificially fragment the cluster rather than identify a genuinely distinct group.

The MST analysis recovers essentially the same groups as the kernel-density analysis of Sec.~\ref{sec:density}, confirming that the structures are robust to the choice of method. The three newly identified groups — Primrose, Rosemary, and Rockrose — are recovered independently by both approaches. The recovered groups are shown in Fig.~\ref{fig:MST}, where the solid white outlines mark the regions defined in Sec.~\ref{sec:stellargroups}.

\subsection{Definition of clusters and groups}
\label{sec:stellargroups}

In this subsection we define the spatial extent of each stellar cluster and group of YSOs in the Rosette Nebula. This will be used in Section \ref{sec:population} to investigate the evolutionary state in various parts of the star forming region.

As a starting point, each stellar group is constrained by a circular geometry, specified by a centre and a radius, introducing no additional free parameters. The radii are chosen to reproduce the reported areas from Table 1 of \citet{Cambresy}, and centred on the coordinates listed in Table 3 of \citet{Roman2}. For groups in close proximity, their regions can overlap. In such cases, the overlapping area is divided along the radical chord. Because this segmentation reduces the effective area of each region, the resulting areas are smaller than the ones reported in the literature. To correct for this, we apply a Monte Carlo–based algorithm that iteratively increases the radii of the circles until the reported areas are reached. For groups without reported areas, plausible values were assigned based on visual inspection. The resulting regions for previously known stellar groups are shown as an overview in Fig.~\ref{fig:RosNebClusters}.

In the following we comment on the defined areas for the main centres of star formation. For more details on how the individual stellar groups were defined, refer to Appendix \ref{clusters}.

\begin{figure}
    \includegraphics[width=\columnwidth]{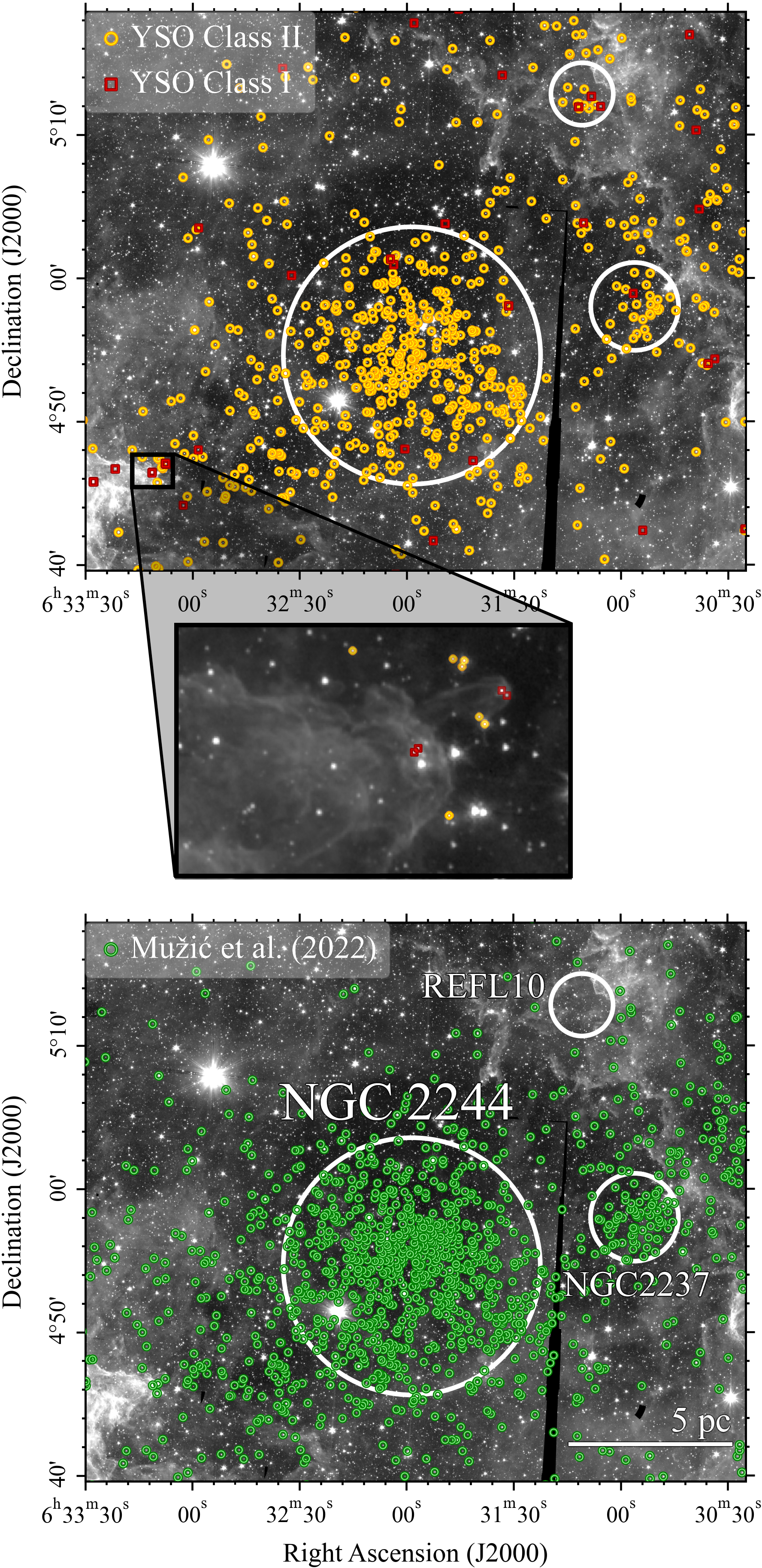}
    \caption{IRAC1 mosaic image of the central cavity of the Rosette Nebula showing the defined regions for NGC~2244, NGC~2237, and REFL10. The upper panel displays Class I (red squares) and Class II (yellow circles) YSO candidates. A zoomed-in view of part of this panel, shown in the middle,  highlights the tip of a giant pillar-like structure. The lower panel shows the probable members of the Rosette Nebula from \citet{Muzic_2022} (green circles). A parsec scale for a distance of 1500~pc is shown in the bottom right.}
    \label{fig:NGC2244}
\end{figure}

\begin{figure*}
    \centering
    \includegraphics[width=\textwidth]{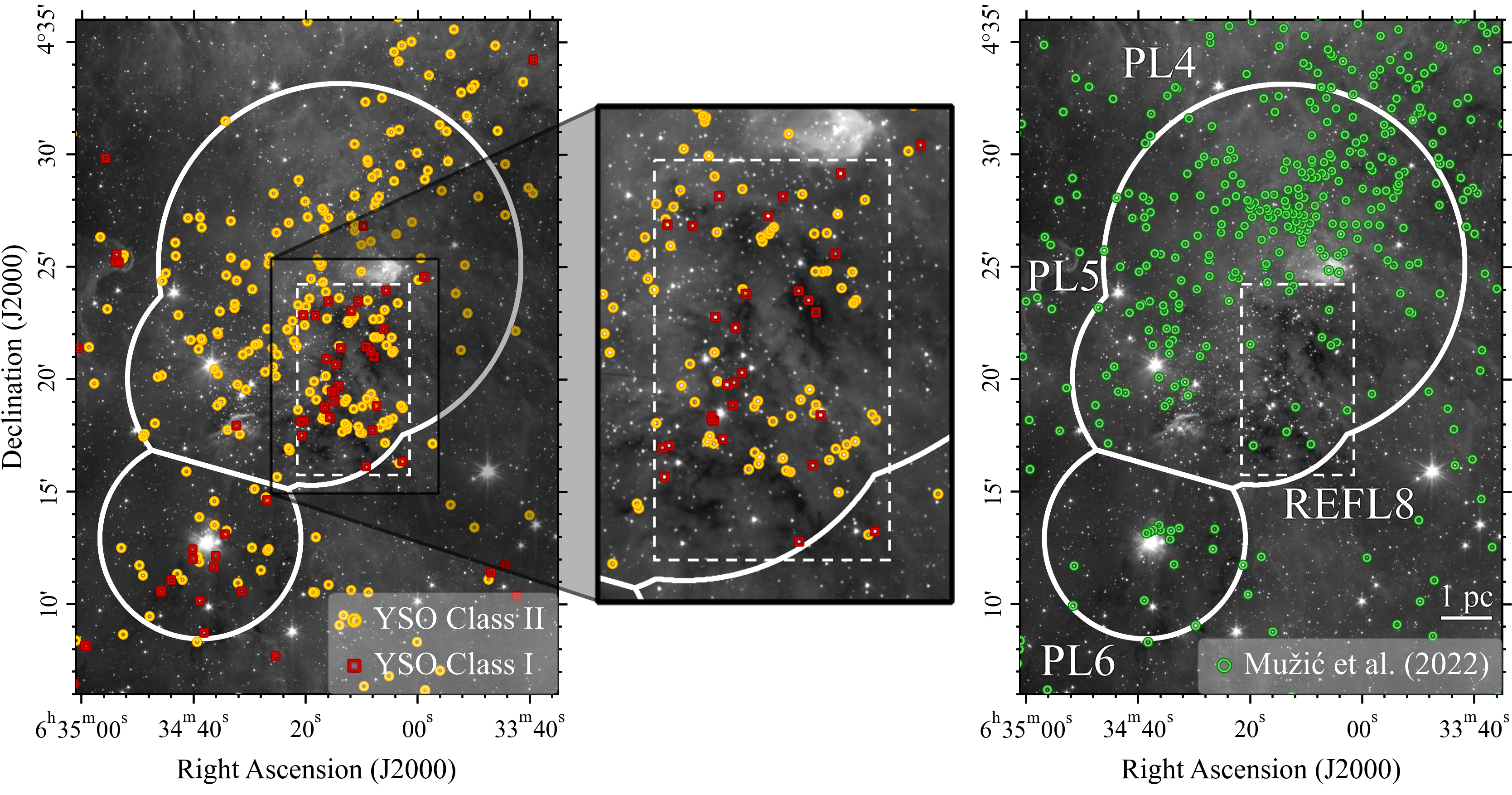}
    \caption{IRAC1 mosaic image of the central core of the Rosette Molecular Cloud showing the defined regions for the PL4–PL5–REFL8 complex and PL6. The left panel displays Class I (red squares) and Class II (orange circles) YSO candidates. A zoomed-in view of part of this panel, shown in the middle, highlights the REFL8 dark cloud (grey dashed rectangle) within the complex. The right panel shows the probable members of the Rosette Nebula from \citet{Muzic_2022} (green circles). A parsec scale for a distance of 1500~pc is shown in the bottom right.}
    \label{fig:RMC_core}
\end{figure*}

In Fig.~\ref{fig:NGC2244}, we display the results for the stellar groups located in the central cavity of the Rosette Nebula, which include NGC~2244, NGC~2237, and REFL10. Further outward, in the Rosette Molecular Cloud, lies another YSO rich environment composed of several stellar groups, shown in Fig. \ref{fig:RMC_core}. These groups are closely spaced on the sky, leading to significant overlap, without any clear demarcation. Therefore, we define a single complex combining PL4, PL5, and REFL8, hereafter referred to as the PL4–PL5–REFL8 complex.

In this complex, there is however a clear segregation between the positions of Class I and Class II YSOs. In Fig.~\ref{fig:RMC_core}, a dark cloud is visible, within which almost all Class I YSOs of the complex are located. This region was approximated by a rectangular area measuring $5'$ by $8.5'$, outlined with a dashed line. This cloud which coincides with a large part of REFL8 is hereafter referred to as the REFL8 dark cloud, while the remainder of the PL4–PL5–REFL8 complex, excluding this cloud, is referred to as PL4–PL5. Note that the combination of PL4–PL5 and REFL8 dark cloud do not exactly match the region of PL4–PL5–REFL8 complex (see Fig.~\ref{fig:RMC_core}).

In Fig.~\ref{fig:MST} the regions defined in this Section are shown as white outlines overlaid on the MST clusters. The two trace each other closely, the adopted boundaries for each stellar group are in good agreement with the groups recovered independently by the MST. Quantitatively, the circular-aperture and MST methods assign a similar number of YSOs to groups -- 773 and 786, respectively -- with an overlap exceeding 90\%. Most of the YSOs missed by the circular-aperture method belong to NGC~2244, largely owing to the adjacent clump that we merge in the MST into the cluster to avoid fragmenting it (Sec.~\ref{sec:mst}). The agreement between two methods with quite different assumptions gives us confidence that the group boundaries adopted here are robust, and we use them in what follows to characterise the evolutionary state of each region.

\section{Structure of star formation}
\label{sec:population}
\subsection{Class I ratio}
\label{sec:CIrat}
In this subsection we estimate the Class I YSO ratio. Since Class I sources are supposed to be at an earlier evolutionary stage, regions with a higher fraction tend to be younger. Therefore, this metric can be used to map star formation activity across the region.

Using the areas defined in Sec.~\ref{sec:stellargroups}, we count the number of Class I and Class II YSOs in each stellar group. The Class I ratio is defined as the fraction between Class I sources and the total number of YSOs (Class I and Class II). These values are reported in Table~\ref{tab:StellarGroupsResults}, ordered by distance from NGC~2244. The Class I ratio uncertainty is computed from binomial statistics using a Beta distribution with a Jeffreys prior. The quoted error bars correspond to the 16th and 84th percentiles of this distribution. For groups with no Class~I detections, we instead report the 84th-percentile of the same distribution as an upper limit on the ratio. The table also reports the area and extinction associated with each region, with extinction values taken from Table 1 of \citet{Cambresy} or estimated using Fig.~2 from the same study.

\setlength{\dashlinedash}{0.2pt} 
\setlength{\dashlinegap}{1.0pt}  
\renewcommand{\arraystretch}{1.2}
\begin{table*}
    \caption{Number of Class I and Class II YSOs per stellar group, along with their Class I ratios.}
    {
    \centering
    \begin{tabular}{l l c c c c}
        \toprule[1pt]
        Stellar Group & Area ($\text{arcmin}^2$) & A$_V$ (mag) & Class I & Class II & Class I ratio (\%)\\
        \midrule[1pt]
        NGC2244         & 254.6 & 2.2  & 5  & 313 & $1.6^{+0.9}_{-0.6}$ \\
        NGC2237         & 29.7  & 2.6  & 1  & 27  & $3.6^{+5.3}_{-2.1}$ \\
        REFL10          & 15.0  & 2.4  & 3  & 9   & $25.0^{+14.0}_{-10.1}$ \\
        PL2             & 13.5  & 8.5  & 1  & 6   & $14.3^{+17.6}_{-8.2}$ \\
        PouC            & 13.5  & 8.5  & 3  & 10  & $23.1^{+13.3}_{-9.3}$ \\
        Primrose        & 10    & 4.0  & 5  & 7   & $41.7^{+14.3}_{-12.9}$ \\
        PL1             & 26.5  & 10.9 & 1  & 11  & $8.3^{+11.4}_{-4.8}$ \\
        PL4-PL5-REFL8 complex &238.4  & 10.2 & 30 & 170 & $15.0^{+2.7}_{-2.3}$  \\
        PouD            & 12.3  & 8.0  & 1  & 17  & $5.6^{+8.0}_{-3.2}$ \\
        PL6             & 57.5  & 12.1 & 12 & 20  & $37.5^{+8.8}_{-8.0}$  \\ 
        PL3             & 39.4  & 10.1 & 1  & 8   & $11.1^{+14.5}_{-6.4}$  \\
        Rockrose        & 15    & 12.0 & 0  & 9   & $< 10.1$ \\
        Rosemary        & 10    & 4.0  & 0  & 13  & $< 7.2$  \\
        PL7             & 19.2  & 10.8 & 15 & 30  & $33.3^{+7.3}_{-6.6}$  \\
        REFL9           & 39.7  & 12.7 & 4  & 41  & $8.9^{+5.1}_{-3.3}$   \\
        \hdashline
        Rosette cavity  &  2100  & --  & 32  & 634   & $4.8^{+0.9}_{-0.8}$ \\
        Interaction ring&  2100  & --  & 50  & 117   & $29.9^{+3.6}_{-3.4}$ \\
        \hdashline
        PL4-PL5        & 197.7  & 8.8  & 3  & 112  & $2.6^{+1.9}_{-1.1}$ \\
        REFL8 dark cloud  & 42.5  & 16.0  & 28  & 59  & $32.2^{+5.2}_{-4.8}$ \\
        \bottomrule[1pt]
    \end{tabular}
    }
    \label{tab:StellarGroupsResults}
\end{table*}

We note that, had we instead used the group memberships from the MST analysis of Sec.~\ref{sec:mst}, the resulting Class~I ratios would be consistent with those reported here, agreeing within their respective uncertainties.

In the same table, below the dotted line, we also include the Rosette cavity and the Interaction ring, which correspond to the regions shown in Fig.~\ref{fig:Result_distribution}. The Rosette cavity refers to the inner dotted circle, while the Interaction ring corresponds to the annular region between the two dotted circles. These circles are not centred exactly on NGC~2244, but are instead centred at $\alpha = 06{:}31{:}41.0$, $\delta = +04{:}52{:}00$, and extend to approximately 11~pc and 16~pc in radius. The inner and outer circles were defined such that the Rosette cavity has the same area as the Interaction ring, with the latter approximately tracing the location of the expanding HII front. Extinction values are not reported for these two regions, as they are spatially extensive and exhibit a wide range of extinction.

The table also includes the regions denoted as PL4–PL5 and the REFL8 dark cloud, which are described in detail in Sec.~\ref{sec:stellargroups}. These regions represent an approximate subdivision of the PL4–PL5–REFL8 complex based on its visible structural features.

\begin{figure}
    \centering
    \includegraphics[width=\columnwidth]{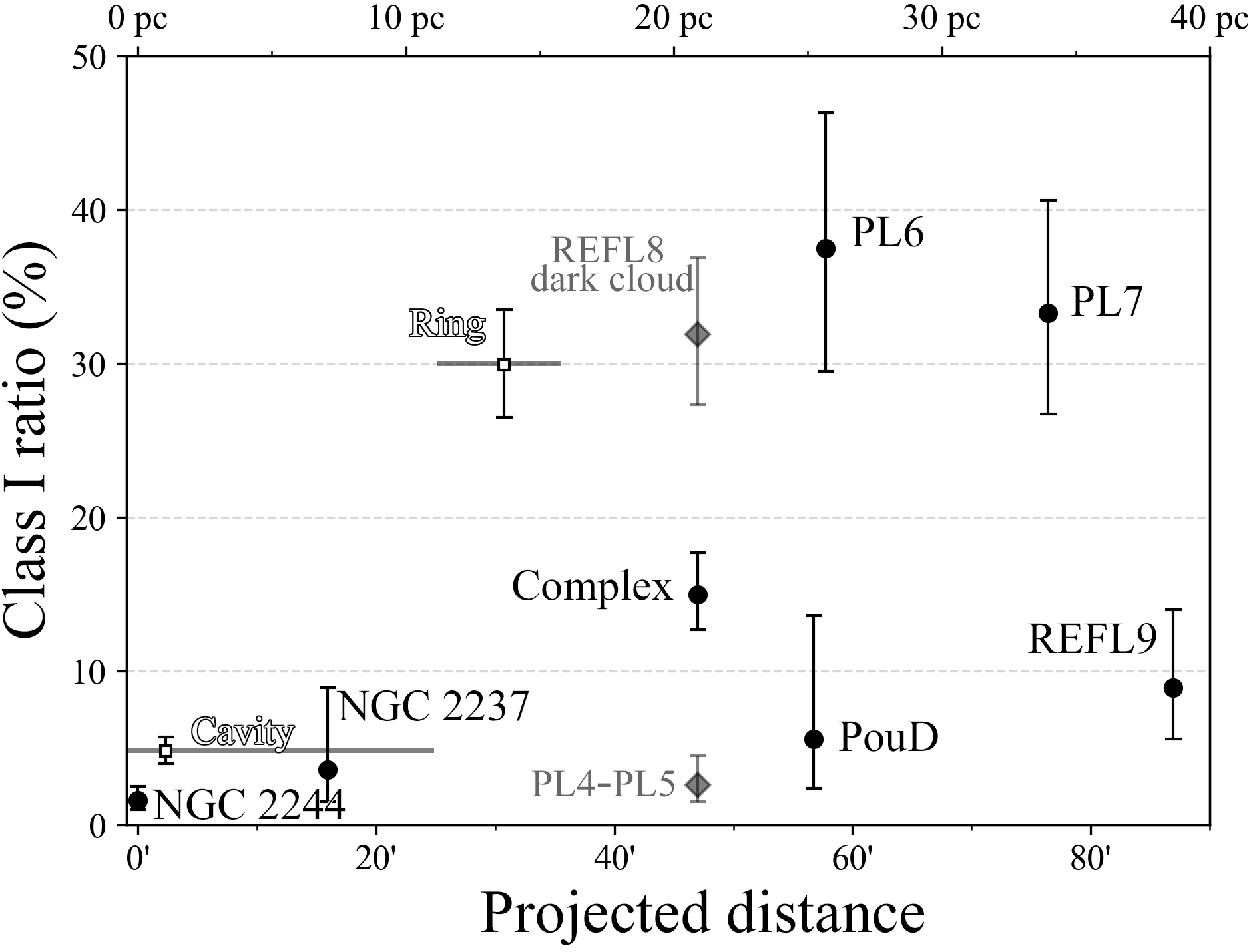}
    \caption{Class I ratio as a function of projected distance from the centre of NGC~2244, expressed in parsecs on the upper x-axis and arcminutes on the lower x-axis. Young stellar groups are shown as black circles, while the subregions of the complex are shown as grey diamonds. The Rosette cavity and the Interaction ring are shown as white squares, with a horizontal line indicating their spatial extent.}
    \label{fig:Class_Frac}
\end{figure}

The Class~I ratios are also shown in Fig.~\ref{fig:Class_Frac}. Regions with uncertainties below 10\% are displayed as a function of their projected distance from NGC~2244, expressed in parsecs on the upper x-axis and arcminutes on the lower x-axis. Although projected distances can be a limiting way to perceive the true distances between stellar groups, most of the molecular cloud lies at nearly the same distance as NGC~2244, with the exception of NGC~2237, which sits nearly 100 pc behind it \citep{Muzic_2022}.

By looking at Fig.~\ref{fig:Class_Frac} it is easy to tell that the Rosette Nebula cavity region, mainly populated by the NGC~2244 and NGC~2237 star clusters, presents low Class I ratios ($<5\%$). On the other hand, the regions where the HII front interacts with the Rosette Molecular Cloud present a higher ratio, around 30\%. More towards the centre of the Rosette Molecular Cloud we see a more mixed situation, including regions with very high fractions like the REFL8 dark cloud and PL6 and others with barely any Class I sources. Towards the far side of the Cloud, we also have a drastic difference between the high fraction in the PL7 stellar group (33\%) and the low fraction from REFL9 (9\%). Overall, the subregions of the Rosette Nebula feature a wide variety in Class I ratios, indicative of its dynamic evolution and several episodes of star formation.

We acknowledge that these Class I ratios could be artificially higher due to spatially varying extinction between groups, where highly reddened Class II sources could mimic the colours of Class I YSOs. However, even regions with higher extinction can show low Class I ratios, such as PL4-PL5 and PouD. Furthermore, Class I YSOs are themselves expected to have high extinctions, so this issue is unlikely to play a major role in the observed differences in Class I ratios.

\subsection{Disk fraction}
\label{sec:discf}
Disk fractions, like the Class I ratios, can be used as indicators of the relative age and evolution of star-forming regions. For this reason, we derive disk fraction values for several stellar groups in the Rosette Nebula.

The disk fraction is calculated as the ratio between the number of sources with disks (Class I and Class II YSOs from this work) and the total number of young stars, including diskless objects (Class I, Class II, and Class III YSOs).
Here, sources are considered disk-bearing only if they show excess emission between 3 and 8$\,\mu m$. Class III YSOs are taken from the catalogue of probable Rosette Nebula members presented by \citet{Muzic_2022}, which we adopt as our diskless population across the region.

We count YSO candidates identified in this work together with the probable members from \citet{Muzic_2022}. When a source is present in both catalogues, it is classified as a disk-bearing object. In addition, only sources with $J$-band magnitudes brighter than 16.5 are considered, corresponding to the completeness limit of the \citet{Muzic_2022} catalogue. This selection criterion reduces our YSO sample to 750 sources and the \citet{Muzic_2022} sample from 2974 to 2745 objects.

The resulting source counts, including disk-bearing YSOs from this work and Class III objects from \citet{Muzic_2022}, are reported in Table~\ref{tab:StellarGroupsResults2}, together with the derived disk fractions. Uncertainties are computed using standard error propagation using Poisson statistics for the numbers of sources with and without disks. Disk fractions are reported only for regions containing at least 15 sources in total, including both disk-bearing and diskless objects. Below this threshold, the large uncertainties in disk fractions render the result not useful.

\setlength{\dashlinedash}{0.2pt}
\setlength{\dashlinegap}{1.0pt}
\renewcommand{\arraystretch}{1.2}
\begin{table*}
    \caption{Number of YSOs with and without disks per stellar group, along with their disk fractions.}
    \centering
    \begin{tabular}{l c c c c}
        \toprule[1pt]
        Stellar Group & A$_V$ (mag) & Class I \& II YSOs & Class III YSOs & Disk fraction (\%)\\
        \midrule[1pt]
        NGC2244         & 2.2  & 215 & 482 & $30.8^{+1.8}_{-1.7}$  \\
        NGC2237         & 2.6  & 22  & 50  & $30.6^{+5.6}_{-5.1}$  \\
        REFL10          & 2.4  & 3   & 0   & -- \\
        PL2             & 8.5  & 3   & 8   & -- \\
        PouC            & 8.5  & 9   & 10  & $47.4^{+11.2}_{-11.0}$ \\
        Primrose        & 4.0  & 3   & 2   & -- \\
        PL1             & 10.9 & 4   & 3   & -- \\
        PL4-PL5-REFL8   & 10.2 & 95  & 170 & $35.8^{+3.0}_{-2.9}$  \\
        PouD            & 8.0  & 6   & 2   & -- \\
        PL6             & 12.1 & 13  & 10  & $56.5^{+9.8}_{-10.3}$  \\
        PL3             & 10.1 & 3   & 8   & -- \\
        Rockrose        & 12.0 & 5   & 0   & -- \\
        Rosemary        & 4.0  & 0   & 0   & -- \\
        PL7             & 10.8 & 16  & 1   & $94.1^{+3.4}_{-8.4}$  \\
        REFL9           & 12.7 & 7   & 3   & -- \\
        \hdashline
        Rosette cavity   & --  & 444 & 948  & $31.9^{+1.3}_{-1.2}$ \\
        Interaction ring & --  & 69  & 171  & $28.8^{+3.0}_{-2.8}$ \\
        \hdashline
        PL4-PL5         & 8.8  & 80  & 162  & $33.1^{+3.1}_{-2.9}$ \\
        REFL8 dark cloud& 16.0 & 15  & 8    & $65.2^{+9.0}_{-10.3}$ \\
        \bottomrule[1pt]
    \end{tabular}
    \label{tab:StellarGroupsResults2}
\end{table*}

\begin{figure}
    \centering
    \includegraphics[width=\columnwidth]{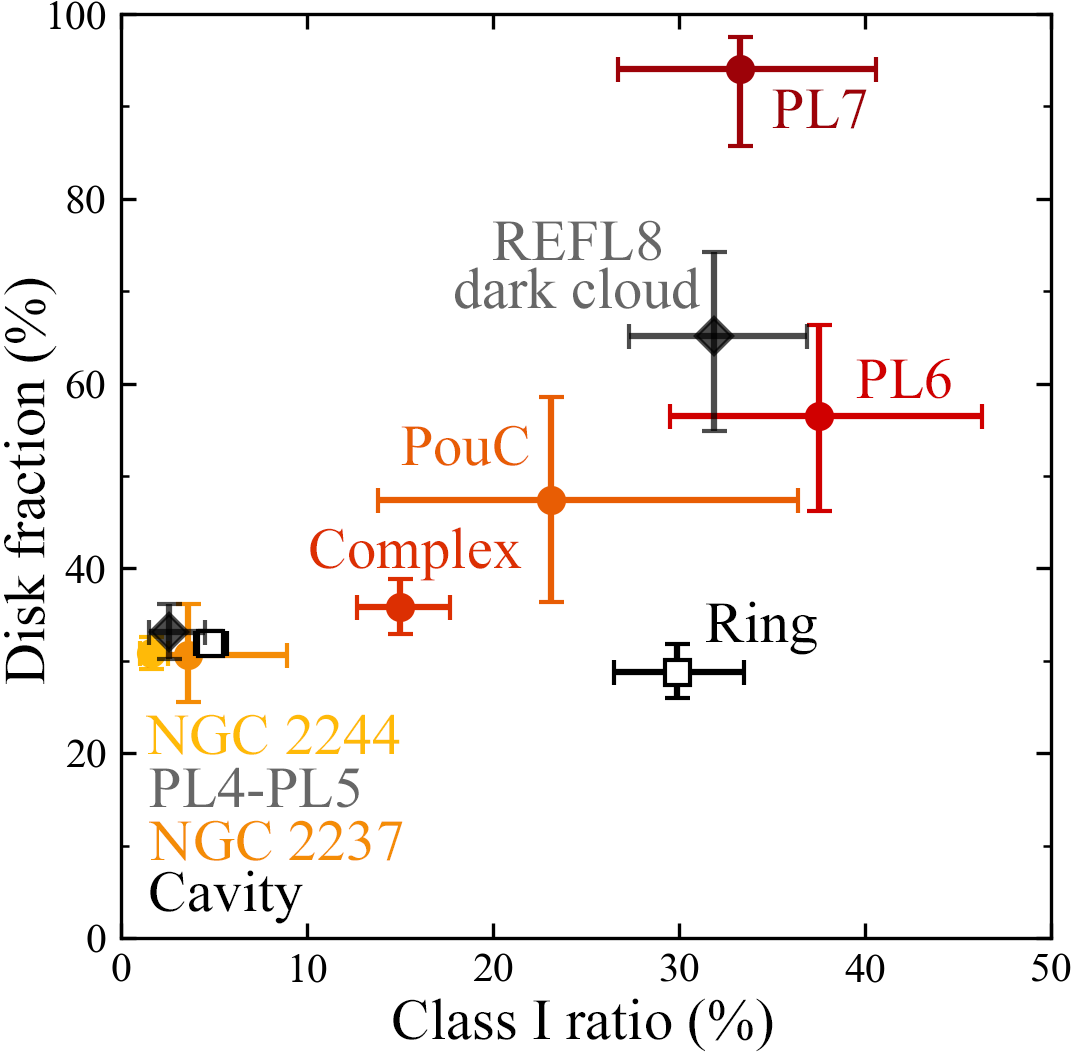}
    \caption{Disk fraction plotted against the Class I ratio. Stellar groups are shown as circles, coloured from yellow to deep red according to their projected distance from NGC~2244. The Rosette cavity and the Interaction ring are represented by white squares, while the PL4–PL5 and REFL8 dark cloud regions are shown as grey diamonds. Uncertainties are indicated for all regions.}
    \label{fig:comp12}
\end{figure}

For regions and stellar groups with reported disk fractions, these values are shown as a function of the Class I ratio in Fig.~\ref{fig:comp12}. We find that the disk fractions of NGC~2244, NGC~2237, PL4–PL5, and the Rosette Nebula cavity are consistent with each other within the uncertainties. Their Class I ratios are similarly consistent. We note that the Interaction ring has a disk fraction similar to the mentioned regions, but a higher Class I ratio.

These regions show disk fractions of 30-35\%, which corresponds to ages of approximately 2-3 Myr \citep{Tatiana}. This is largely consistent with the age estimates of $1.3 \pm 0.4$ Myr for NGC~2244 and $1.6 \pm 0.5$ Myr for the Rosette Nebula as a whole reported by \citet{Muzic_2022}. Studies of NGC~2244 using MIR data report disk fraction values of 45\% \citep{Balog} and $39\pm9$\% for sources with a spectral type later than K0 \citep{Almendros2023}. Our value of $31\pm2$\% for this cluster is a bit lower but still within the combined error bars.

In contrast, the REFL8 dark cloud, as well as PL6 and PL7, exhibit significantly higher disk fractions, ranging from about 55\% to 95\%, albeit with larger uncertainties. Such high disk fractions are indicative of younger ages, closer to 1 Myr or possibly younger \citep{Tatiana}. Consistent with this interpretation, these regions also show the highest Class I ratios, as expected for younger stellar populations. Overall, among the stellar groups shown in Fig.~\ref{fig:comp12}, PouC, PL6, PL7, REFL8, and the Interaction ring stand out as the most actively star-forming regions, while the remaining groups appear to be largely coeval.

As with the Class I ratio analysis, disk fractions can be affected by extinction differences between stellar groups. However, as with the Class I ratios, we find a wide range of disk fractions even among groups with high extinction ($A_V > 8$~mag), indicating that this is not an obvious bias.

\subsection{Star formation chronology}
\label{sec:chronology}

The Class I ratios and disk fractions derived in Sec.~\ref{sec:CIrat} and \ref{sec:discf} are useful indicators of the evolutionary stage of the stellar groups, and in the following we use them to reconstruct the chronology of star formation in the Rosette Nebula. However, we would like to caution against overinterpreting these numbers, as they can also be affected by variations in extinction and sensitivity across the region, and in some cases the sample sizes are too small for robust statements.

The Rosette Nebula cavity hosts some of the oldest populations among the identified young stellar groups. Star formation appears to have begun at a similar epoch at NGC~2244, NGC~2237, and in parts of the molecular cloud such as PL4-PL5 and PouD. At the outer edges, REFL9 also appears to have formed around this time, or shortly thereafter. These groups all show Class I ratios $\lesssim$5\%, with REFL9 closer to 10\%. The disk fractions available for some of these clusters and groups point to the same conclusion, with values around 30\%, corresponding to an age of a few Myr.

On the other hand, the youngest stellar groups -- with the highest Class I ratios (>30\%) and disk fractions (>50\%) -- are PL6, PL7, and the REFL8 dark cloud, all located beyond the Interaction ring. Their disk fractions suggest ages of 1~Myr or younger. In PL6 and REFL8, the presence of prominent dark clouds in the IRAC mosaics (see Fig.~\ref{fig:RMC_core}) suggests that star formation is likely still ongoing.

We note that some of these groups lie at large projected distances from NGC~2244 -- REFL9 in particular, at nearly 40~pc -- making a sequential triggering scenario driven by compression from the expanding HII region implausible. Indeed, there is no evidence that the onset of star formation in these groups is correlated.

A few groups have evolutionary metrics that indicate an age in between the youngest regions and NGC~2244. These have Class I ratios between 10 and 30\% and disk fractions between 30 and 50\%. These groups sit near the HII front, along the Interaction ring, a region that hosts a large population of YSOs together with groups such as PL1, PL2, PouC, and Primrose. Triggered star formation offers a viable explanation for these intermediate values, consistent with star formation occurring after the formation of NGC~2244: the expanding HII bubble may facilitate compression of the surrounding gas via the strong stellar winds from NGC~2244's OB population, potentially promoting star formation in the Rosette Molecular Cloud. This scenario has also been supported by previous studies \citep{Poulton, Roman2, Cambresy}.

While star formation in some groups near the cavity may have been aided by the expansion of the HII region driven by NGC~2244, this mechanism cannot account for the full spatial and temporal distribution of the young stellar populations. We reiterate that the relative age sequence found does not support a simple scenario of sequential triggered star formation across the Rosette Nebula. Instead, it points to a more intricate and dynamic evolution of the molecular cloud. These conclusions are consistent with previous studies \citep{PL7, Poulton, Schneider, Ybarra, Cambresy, Lim, Muzic_2022}.

\section{Summary}
\label{sec:conclusion}

In this paper we present a deep infrared survey of the Rosette Nebula, based on stacked Spitzer/IRAC images. While parts of the underlying dataset have been published before, this is the first combined analysis of all available Spitzer/IRAC data for this region. We compile a catalogue of $>300000$ sources out of which close to 20000 have valid photometry from 3 to 8$\,\mu m$. The catalogue includes the majority of the cloud, with 2.7 square degrees being covered in all four IRAC channels. 

We use a set of colour-colour criteria to classify the sources in our multi-band catalogue, largely following the classification scheme by \citet{Gutermuth}. Overall, we classify 1528 sources as YSOs, 1303 are Class II, and 225 Class I. About half of this sample is new and has not been identified before as YSOs in the literature. In comparison with literature samples, our YSO census extends to fainter magnitudes. It is about 1\,mag deeper in J-band than the recent optical survey by \citet{Muzic_2022}. In terms of masses, our sample extends into the substellar domain and includes brown dwarf candidates in this region.

Based on our large sample of YSOs, we analyse the star formation activity in the Rosette Nebula. The distribution of Class II and Class I YSOs is highly inhomogeneous, with a large number of stellar clusters and groups. We confirm almost all known young clusters of YSOs in this region, and identify three new groups, each with around 10 YSOs in an area of 10-15 arcmin$^2$. We define the area for each cluster using the literature, geometric criteria, and an MST algorithm. More than 600 YSOs are found in the central cavity of the nebula, about half of them in the well known cluster NGC~2244. Close to 150 further YSOs are found in an annulus around the cavity with a radius of 11-16\,pc, which contains the previously known regions PL1, PL2, and PouC. This region approximately coincides with the HII ionisation front in the nebula. Another active region of star formation is a complex combining groups previously named PL4, PL5, and REFL8, with about 200 YSOs in total, which is located beyond the annulus. 

For each grouping of YSOs, we calculate the ratio between Class I and Class II sources. For some clusters we also derive a disk fraction, by combining our sample with that of \citet{Muzic_2022}. Both Class I ratio and disk fraction are used here as an indicator of the evolutionary state of the underlying sample. Clusters in the cavity have Class I ratios below 5\% and disk fractions around 30\%, indicating that star formation activity has ceased and a typical age of a few Myr. Other clusters scattered around the nebula share similar characteristics, for example, PL4 and PL5. The previously mentioned annulus has a similar disk fraction, but a significantly higher Class I ratio of 30\%, indicating that this ring harbours active centres of star formation. High Class I ratios (>30\%) as well as disk fractions (>50\%) are found in the clusters PL6, PL7, as well as in the dark cloud region that mostly coincides with a cluster formerly known as REFL8. These three are roughly at 20\,pc distance from the centre, past the ionization front.

Overall, our results confirm the previously established view of the star formation chronology in the Rosette Nebula (see Sec.~\ref{sec:chronology}). An initial wave of primordial star formation happened about 2\,Myr ago and formed clusters in the centre and in several other places in the nebula. There is evidence for ongoing activity in a ring around the central cavity; here star formation is likely being triggered by the HII expansion. In addition, there are active centres of star formation outside the ionisation front. Taken together, this points to multiple pathways to star formation within the molecular cloud, which results in a complex, multi-stage star forming history.

\section*{Acknowledgements}
D.C. acknowledges support from FCT - Fundação para a Ciência e a Tecnologia, I.P. through the individual grant 2025.02513.BD (\url{https://doi.org/10.54499/2025.02513.BD}). A.S. and B.D. acknowledge support from the UKRI Science and Technology Facilities Council through grant ST/Y001419/1/. K.M. acknowledges support from the Fundação para a Ciência e a Tecnologia (FCT) through the CEEC-individual contract 2022.03809.CEECIND, and grant UID/04434/2023.
The authors declare that in the preparation of this paper LLM tools were used, in particular GPT-5, and Claude Opus 4.8/5. Specifically, these tools were used to produce and refine limited parts of the code used in the analysis, under specific and strict guidance by the authors. The authors confirm that all LLM outputs were verified thoroughly by the authors.

\section*{Data availability}

The data underlying this article are publicly available in the Spitzer Heritage Archive; the UKIDSS Galactic Plane Survey database; the 2MASS All-Sky Point Source Catalog; the Pan-STARRS1 Surveys data archive; and the Gaia Archive. The YSO catalogue produced in this work will be made available via VizieR upon publication.



\bibliographystyle{mnras}
\bibliography{references} 



\appendix

\section{Definition of clusters and groups}
\label{clusters}

In the following we describe in detail the definition of clusters and groups of YSOs in the Rosette region.

In Fig.~\ref{fig:NGC2244} we show the defined regions for NGC~2244, together with those for NGC~2237 and REFL10. For NGC~2244, the area reported by \citet{Cambresy} is fragmented, as illustrated in Fig. 6 of that work. We therefore adopted an area 50\% larger to encompass a larger fraction of the cluster members.

Historically, the designation NGC~2237 has been used both to describe part of the nebulous region and, at times, the entire Rosette Nebula. More recent studies, however, use NGC~2237 to refer specifically to a cluster located west of NGC~2244 (see Fig.~\ref{fig:NGC2244}). The coordinates adopted for the centre of NGC~2237 are offset by approximately $5'$ from the reported values. We therefore centre NGC~2237 at $\alpha = 06{:}30{:}56.1$, $\delta = +04{:}58{:}00.6$.

For REFL 10, located within the cavity north of NGC~2237 (see Fig.~\ref{fig:NGC2244}), the central coordinates were offset by approximately $2'$ from the one reported in \citet{Roman2} to $\alpha = 06{:}31{:}10.79$, $\delta = +05{:}12{:}50$. Also based on a visual inspection, an area of 15 $\mathrm{arcmin}^2$ was adopted, as no value is reported by \citet{Cambresy}.

\begin{figure*}
    \centering
    \includegraphics[width=\textwidth]{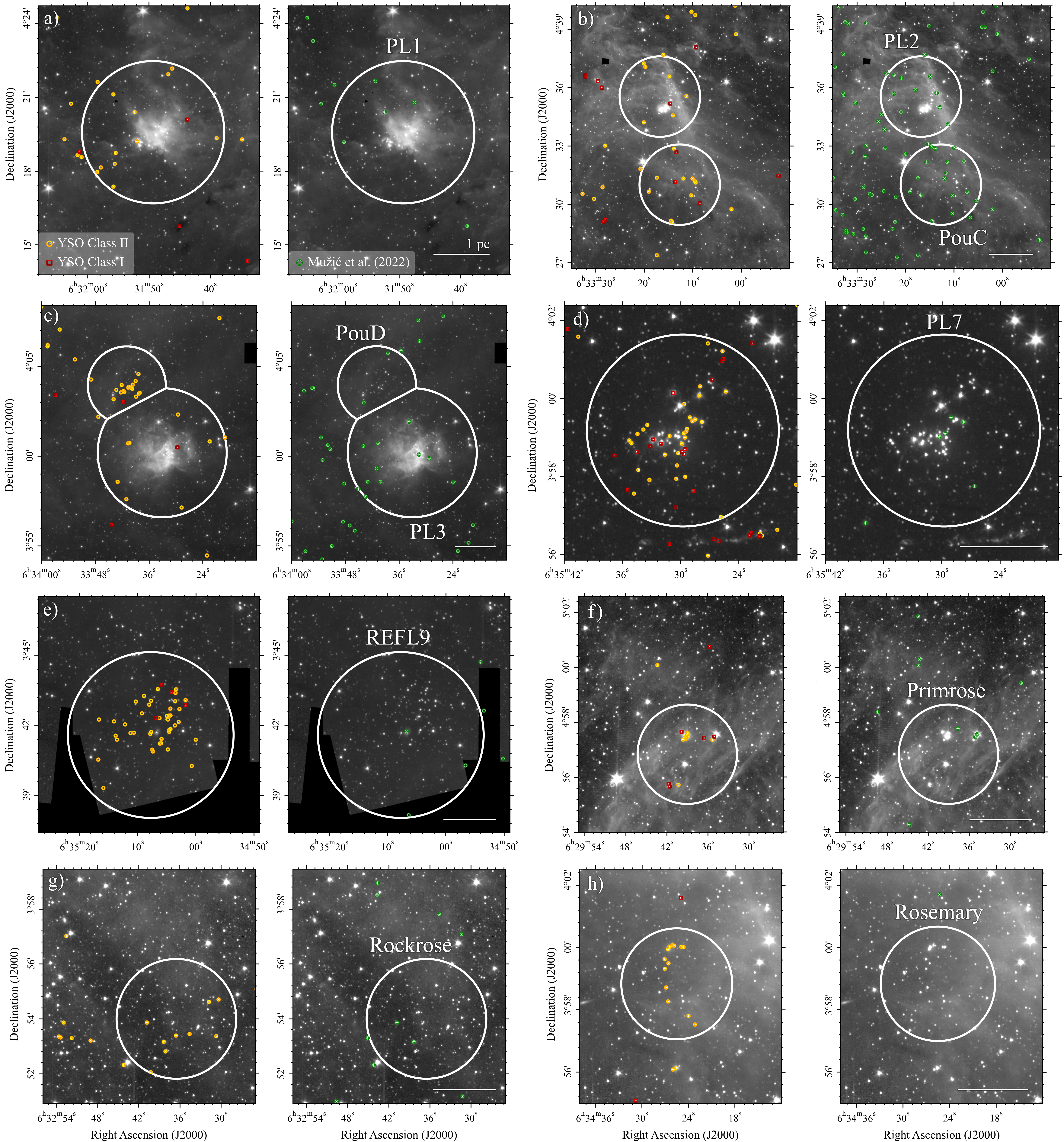}
    \caption{IRAC1 mosaic images of the young stellar groups highlighting their defined regions (with a thick white outline for previously known groups and a thick white dashed circle for newly discovered stellar groups). Each stellar group is presented in a pair of panels: \textit{a)} PL1, \textit{b)} PL2 (north) and PouC (south), \textit{c)} PouD (north) and PL3 (south), \textit{d)} PL7, \textit{e)} REFL9, \textit{f)} Primrose, \textit{g)} Rockrose, and \textit{h)} Rosemary. The left panel of each pair shows Class I (red squares) and Class II (orange circles) YSO candidates, while the right panel displays the probable members of the Rosette Nebula from \citet{Muzic_2022} (green circles). A one parsec scale for a distance of 1500~pc is shown in the bottom right for each pair of panels.}
    \label{fig:AllG}
\end{figure*}

In the Interaction ring between the expanding HII and the cold molecular cloud we find PL1, PL2 and PouC, shown in Fig.~\ref{fig:AllG} \textit{a)} and \textit{b)}. For PouC (to the south of PL2, see Fig.~\ref{fig:AllG} \textit{b)}), no area is reported by \citet{Cambresy}, and we therefore assume an area equal to that of PL2.

In the centre of the Rosette Molecular Cloud, shown in Fig. \ref{fig:RMC_core}, are several stellar groups that are very closely spaced, leading to significant overlap. Here we use an algorithm to iteratively increase the radii of the regions by equal increments until the total cumulative area matches the reported values. However, to prevent the PL5 region from extending into the PL6 area, the radius of PL5 is fixed from the start and therefore kept unchanged. The final area assigned to the PL4–PL5–REFL8 complex corresponds to the sum of the areas reported for each group in the literature.

Simultaneously, for PL6, which lies immediately south of the complex (see Fig. \ref{fig:RMC_core}), the boundary is defined where the PL4–PL5–REFL8 complex meets the circular geometry of PL6, resulting in a segmented circular shape. Despite this non-circular geometry, the literature area is preserved. The final configurations are shown in Fig.~\ref{fig:RMC_core}.

The PL3 and PouD stellar groups are also closely spaced on the sky (see Fig.~\ref{fig:AllG} \textit{c)}). Although \citet{Cambresy} does not report an area for PouD, Fig. 6 of that work shows that the area reported for PL3 encompasses PouD. Based on a visual inspection of our YSO candidates, we assigned part of the reported area to PL3 and the remainder to PouD.

Finally, at the edge of the main part of the cloud we have PL7 (see Fig.~\ref{fig:AllG} \textit{d)}) and REFL9 (see Fig.~\ref{fig:AllG} \textit{e)}). These were straightforward cases, for which we simply adopted the areas reported in the literature.

The newly identified stellar groups, Primrose, Rockrose, and Rosemary, and their corresponding defined regions are shown in Fig.~\ref{fig:AllG} \textit{f)}, \textit{g)} and \textit{h)}. For Primrose, an area of 10 arcmin$^{2}$ was adopted based on visual inspection, centred at $\alpha = 06{:}29{:}39.0$, $\delta = +04{:}56{:}50.0$. For Rockrose, an area of 15 arcmin$^{2}$ was chosen, centred at $\alpha = 06{:}32{:}36.5$, $\delta = +03{:}54{:}00.0$. For Rosemary, an area of 10 arcmin$^{2}$ was adopted, centred at $\alpha = 06{:}34{:}25.5$, $\delta = +03{:}58{:}50.0$.

All YSOs located within these defined regions are considered to be members of that associated stellar group. Of the 225 Class I YSOs, 143 lie outside these regions, while 612 out of 1303 Class II YSOs are also located outside. In the \citet{Muzic_2022} sample, which contains 2974 sources, only around one third fall within the defined stellar group regions, leaving 1948 outside.

\bsp	
\label{lastpage}
\end{document}